\documentclass[preprint,aps,12pt,showpacs,nofootinbib,tightenlines]{revtex4}
\usepackage{mathrsfs}
\usepackage{amsmath}
\usepackage{amssymb}
\usepackage{epsfig}
\usepackage{epstopdf}
\usepackage{graphicx}
\usepackage{subfigure}
\usepackage{booktabs}
\usepackage{float}
\usepackage{amsmath}
\usepackage{color}
\usepackage{multirow}
\usepackage{bm}

\def\be{\begin{eqnarray}}
\def\en{\end{eqnarray}}
\def\non{\nonumber\\}

\begin{document}
\title{$D^*_{(s)}\to P$ form factors and their applications to semi-leptonic and non-leptonic weak decays}
\author{Hao Yang$^1$, Zhi-Qing Zhang$^1$
\footnote{zhangzhiqing@haut.edu.cn},  You-Ya Yang$^1$, Peng Li$^2$\footnote{lipeng@haut.edu.cn} } 
\affiliation{\it \small $^1$  School of Physics, Henan University
of Technology, Zhengzhou, Henan 450052, China\\
\it \small $^2$ Institute for Complexity Science, Henan University of Technology,
Zhengzhou, Henan 450001, China } 
\date{\today}
\begin{abstract}
Similar to other heavy flavor mesons, the weak decays of $D^*_{(s)}$ mesons can provide a platform to verify the standard model, explore new physics, and understand the mechanisms of weak interactions. At present, the theoretical and experimental studies on $D^*_{(s)}$ mesons are relatively limited. In addition to the dominant electromagnetic decays, the $D^*_{(s)}$ weak decays should be feasible to explore the $D^*_{(s)}$ mesons. In this study, we used the covariant light-front quark model to study the form factors of the transitions $D^*_{(s)}\to \pi, K, \eta_{q,s}$, then calculated the branching ratios of the semi-leptonic decays $D^*_{(s)}\to P\ell^{+}\nu_{\ell}$ and the non-leptonic decays $D^*_{(s)}\to PP, PV$ with $P=\pi, K, \eta^{(\prime)}, V=\rho, K^*, \phi$ and $\ell=e, \mu$. The channels $D_{s}^{*+}\to\eta \ell^{+}\nu_{\ell}$ and $D^{*+}_{s}\to \eta\rho^{+}$ possess the largest branching ratios, which can reach an order of $10^{-6}$ among these decays, and are most likely to be accessible in experiments at future high-luminosity colliders.
Furthermore, we predict and discuss the longitudinal polarization fraction $f_{L}$ and the forward-backward asymmetry $A_{FB}$ for the considered
semi-leptonic $D^*_{(s)}$ decays.
\end{abstract}

\pacs{13.25.Hw, 12.38.Bx, 14.40.Nd} \vspace{1cm}

\maketitle

\section{Introduction}\label{intro}
In the 1970s and 1980s, Terentev and Berestesky proposed the light-front quark model (LFQM) \cite{Terentev,Berestetsky}, which aims to deal with non-perturbable physical quantities such as decay constants and transition form factors \cite{AJaus,BJaus,AChoi}. However, the standard LFQM has trouble dealing with zero-mode contributions. To address this limitation, Jaus developed an improved model, the covariant light-front quark model (CLFQM)\cite{CJaus}. Compared with other quark-model approaches, the CLFQM has some unique advantages. In this approach, the light-front wave functions describing the hadron through quark and gluon degrees of freedom can preserve a Lorentz
invariant formalism. As the final state meson at the maximum recoil point $q^2=0$ is usually relativistic, considering the relativistic effects is necessary. From this perspective, one can expect that the relativistic CLFQM should be more suitable to study hadronic transition form factors compared with the non-relativistic quark model. Furthermore, the additional spurious contributions apearing in the
previous LFQM are just canceled by the zero-mode contributions under the CLFQM, thereby the result is guaranteed to be covariant. This model has been successfully used in the study of non-leptonic and semi-leptonic meson decays \cite{ZhangSun,yangy,WangShen,LiuKe,ChengHwang,WangLu}.

The weak decays of the $D^{*}_{(s)}$ mesons provide another important platform and opportunity to understand the properties of the $D^{*}_{(s)}$ mesons, explore their decay mechanism and verify the standard model (SM). Owing to the low strength of the weak interactions, the $D^{*}_{(s)}$ weak decays are usually very rare processes. At present, only a few decay modes of $D^*_{(s)}$ mesons have been observed in experiments, that is, $D^*_{(s)}\to D_{(s)}\pi, D_{(s)}\gamma, D_{(s)}e^+e^-$. Very recently, with the advancement of experimental techniques, the BESIII collaboration has reported the first experimental study of the purely leptonic decay $D^{*+}_s \rightarrow e^{+}\nu_e$ \cite{BESIII2023} with the branching ratio  measured as $(2.1^{+1.2}_{-0.9}\pm0.2)\times10^{-5}$. 
Although the theoretical predictions about the $D^{*}_{(s)}$ properties are still relatively limited, and the information regarding the $D^{*}_{(s)}$ weak decays is still very scarce, the experimental progress of the investigation of $D^{*}_{(s)}$ mesons at various high energy collider experiments will provide more data on the $D^{*}_{(s)}$ meson decays, hence the theoretical study of the $D^{*}_{(s)}$ weak decays should have broad prospects in the near future.

 Assuming that the exclusive cross sections near threshold $\sigma(e^+e^- \to D^0 \bar{D}^{*0})\approx \sigma(e^+e^- \to D^+ D^{*-})\approx$ 4nb and $\sigma(e^+e^- \to D^{*0} \bar{D}^{*0})\approx \sigma(e^+e^- \to D^{*+} D^{*-})\approx$ 3nb, more than $5\times 10^7$ $D^{*\pm}$ meson events have been accumulated corresponding to a total integrated luminosity of 15.7 fb$^{-1}$ within the energy region $\sqrt{s} \in [4.085,4.600]$ in BESIII experiments \cite{Ablikim}. Note that the cross-section values considered here are only a rough approximation based
 on the results published by BESIII experiment. However, this is not sufficient to explore the $D^*$ meson weak decays. In the future, approximately $8 \times 10^{10}$ $D^{*0}$ and $D^{*\pm}$ events will be produced at the $\tau$-charm factory (STCF) at a total integrated luminosity of 10 ab$^{-1}$ \cite{STCF}. Given the charm quark fragmentation fractions $f(c\to D^{*+})\approx25\%$ and  $f(c\to D^{*0})\approx23\%$ \cite{Lisovyi}, more than $2\times 10^{10}$ $D^{*0}$ and $D^{*\pm}$ mesons will be collected at SuperKEKB \cite{KEKB}. It is expected that about $10^{12}$ and $10^{13}$ $Z^0$ bosons will be available at the Circular Electron Positron Collider (CEPC) \cite{CEPC} and at the Future Circular Collider (FCC-ee) \cite{FCCee} with a total integrated luminosity 20 ab$^{-1}$, respectively. Considering the branching ratios $\mathcal{B} r(Z^0 \to D^{*0}X/\bar D^{*0})\approx\mathcal{B} r(Z^0 \to D^{*\pm}X)=(11.4\pm1.3)\%$ \cite{Zyla}, more than $10^{11}$ and $10^{12}$ $D^*$ mesons can be obtained at CEPC and FCC-ee, respectively. In addition, with the inclusive cross section $\sigma(pp \to D^{*+}X)=784\pm4\pm87\; \mu b$  at the center of mass energy $\sqrt{s}=13$TeV  measured by the LHCb \cite{Aaij}, more than $2\times 10^{14}$ $D^*$ mesons with a total integrated luminosity of 300fb$^{-1}$ will be collected at the High Luminosity LHC (HL-LHC) experiments up to 2037 \cite{Bediaga}.
 Assuming the exclusive cross sections near threshold $\sigma(e^+e^- \to D^+_s D^{*-}_s)$ and $\sigma(e^+e^- \to D^{*+}_s D^{*-}_s)$ as 1.0 nb and 0.2 nb \cite{Dong,Pakhlova,Sanchez}, approximately $10^{10}$ $D^{*\pm}_s$  events corresponding to a data sample of 10 ab$^{-1}$ will be available at STCF. Given the branching ratio $\mathcal{B}r(Z^0 \to c \bar c)=(12.03\pm0.21)\%$ and the fragmentation fraction $f(c\to D^*_s)\simeq 5.5\%$ \cite{Lisovyi}, approximately $1.3 \times 10^{10}$ and $6.6 \times 10^{10}$ $D^{*\pm}_s$ events corresponding to $10^{12}$ and $5 \times 10^{12}$ Z bosons will be collected at the future CEPC \cite{CEPC} and FCC-ee \cite{FCCee} experiments, respectively. Given the fragmentation fraction $f(c\to D^*_s)\simeq 5.5\%$ \cite{Lisovyi}, approximately $5.5 \times 10^9$ \;$D^*_s$ events corresponding to $5 \times 10^{10}\; c \bar c$ pairs can be collected in the future SuperKEKB experiments \cite{KEKB}. In addition, considering the inclusive cross section $\sigma(pp \to c\bar c X)=2.4$ mb at the center-of-mass energy of $\sqrt{s}=13$ TeV \cite{Aaij} and the charm quark fragmentation fraction $f(c\to D^{*}_s)\approx5.5\%$, about $4\times 10^{13}$ $D^{*}_s$ events corresponding to a data sample of 300 fb$^{-1}$ can be collected at the LHCb \cite{HLLHC}.

In the semi-leptonic decays, the calculations of the hadronic matrix elements are crucial and can be characterized by several form factors \cite{Sun2023}, which can be extracted from data or obtained using some non-perturbative methods. As one of popular non-perturbative approaches, the CLFQM has been successfully used to calculate the form factors \cite{ZhangSun,yangy,ChengChua,C.W.Hwang,Z.T.Wei,ShenWang}. As to the non-leptonic decays involving two hadrons in the final states, the related dynamics becomes more complex when the long distance interactions are involved. For some non-leptonic decays governed by the tree operators, the corresponding matrix elements can be decomposed into the product of the decay constant and the transition form factor by means of the vacuum saturation hypothesis. Such factorization approach is verified to work well for the color-allowed decay modes and has been widely used in the analysis of the non-leptonic weak decays \cite{Sun2023}. In conclusion, the form factors are of great significance for studying the semi-leptonic and non-leptonic weak decays. Various nonperturbative approaches, such as fist-principles lattice QCD \cite{mengyu}, QCD sum rules (QCDSR) \cite{YuMingWang,YuMingWang11}, Bauer-Stech-Wirbel (BSW) model \cite{R. Dhir}, Bethe-Salpeter method \cite{T. Wang}, light-cone sum rules \cite{wangl} and quark models \cite{kang},  have also been used to study the transition form factors. For recent developments in the study of heavy-to-light form factors, one can refer to Refs. \cite{ymwang1,ymwang2}. Combining the form factors with helicity amplitudes, in addition to the branching ratios, we calculate two physical observables, namely the longitudinal polarization fraction $f_{L}$ and forward-backward asymmetry $A_{FB}$. These observables provide valuable insights into the underlying dynamics of the considered decay processes and important constraints on testing the SM.

The remainder of this paper is organized as follows. The formalisms of the CLFQM, hadronic matrix elements and helicity amplitudes combined via form factors are listed in Sec. \ref{form1}. In addition to the numerical results of the form factors of the transitions $D^{*}_{(s)}\to K,\pi,\eta_{q,s}$, the branching ratios, longitudinal polarization fractions $f_L$ and forward-backward asymmetries $A_{FB}$ for the corresponding decays are presented in Sec. \ref{numer}. Comparation with other theoretical results and relevant discussions are also included. The summary is given in Sec. \ref{sum}. In Appendix A and B, some specific rules in the process of performing $p^{-}$ integration as well as expressions for the form factor are presented, respectively.

\section{Formalism}\label{form1}
The form factors for the transitions $D^{*}_{(s)}\to P$ are defined as follows,
	\begin{footnotesize}
		\begin{eqnarray}
			\left\langle P\left(P^{\prime\prime}\right)\left| V_{\mu}\right|D_{(s)}^*\left(P^{\prime}, \varepsilon^*\right)\right\rangle &=&-\frac{1}{m_{D_{(s)}^*}+m_{P}} \epsilon_{\mu \nu \alpha \beta} \varepsilon^{* \nu} P^{\alpha} q^{\beta} V^{D_{(s)}^* P}\left(q^{2}\right),\nonumber\\
			\left\langle P\left(P^{\prime\prime}\right)\left| A_{\mu}\right|D_{(s)}^*\left(P^{\prime}, \varepsilon^*\right)\right\rangle &=& i\left\{\left(m_{D_{(s)}^*}+m_{P}\right) \varepsilon_{\mu}^{*} A_{1}^{D_{(s)}^* P}\left(q^{2}\right)-\frac{\varepsilon^{*} \cdot P}{m_{D_{(s)}^*}+m_{P}} P_{\mu} A_{2}^{D_{(s)}^* P}\left(q^{2}\right)\right. \nonumber\\
			&& \left.-2 m_{P} \frac{\varepsilon^{*} \cdot P}{q^{2}} q_{\mu}\left[A_{3}^{D_{(s)}^* P}\left(q^{2}\right)-A_{0}^{D_{(s)}^* P}\left(q^{2}\right)\right]\right\},
			\label{Dsp}
		\end{eqnarray}
	\end{footnotesize}
where $P=P'+P'', q=P'-P''$, and the longitudinal polarization $\varepsilon(0)=\frac{1}{M_0}(\frac{-M^2_0+P^2_\perp}{P^+}, P^2, P^\perp)$ with $M_0$ is the squared kinetic invariant mass of the meson. In Eq. (\ref{Dsp}), $V_{\mu}$ and $A_{\mu}$ are the vector and axial-vector currents, which are dominant contributions in the weak decays. The four-momentum of the initial (final) meson is $P'=p^\prime_1+p_2\; (P''=p''_1+p_2)$, where $p_1^{\prime(\prime\prime)}$ and $p_2$ are the momenta of the quark and antiquark inside the incoming (outgoing) meson. These momenta can be expressed in terms of the internal variables $(x_i, p^\prime_\perp)$,
\be
p^{\prime+}_{1,2}=x_{1,2} P^{\prime+}, \hspace{0.8cm} p^{\prime}_{1,2\perp}=x_{1,2} P^{\prime}_\perp \pm p^{\prime}_{\perp}
\en
with $x_1+x_2=1$, $x_2=x$. Note that we use $P^\prime = (P^{\prime+}, P^{\prime-}, P^\prime_\perp)$, where $P^{\prime\pm}=P^{\prime0}\pm P^{\prime3}$, so that $P^{\prime2}=P^{\prime+}P^{\prime-}-P^{\prime2}_\perp$. Some internal quantities of on-shells quarks are defined in Appendix A.

 \begin{figure}[htbp]
 	\centering \subfigure{
 		\begin{minipage}{5cm}
 			\centering
 			\includegraphics[width=5cm]{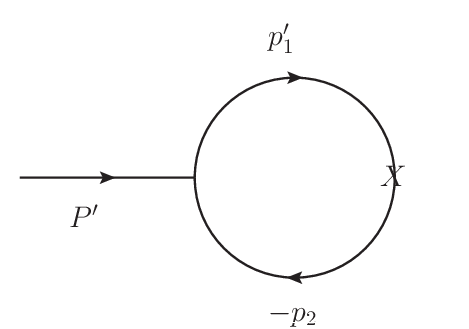}
 	\end{minipage}}
 	\subfigure{
 		\begin{minipage}{6cm}
 			\centering
 			\includegraphics[width=6cm]{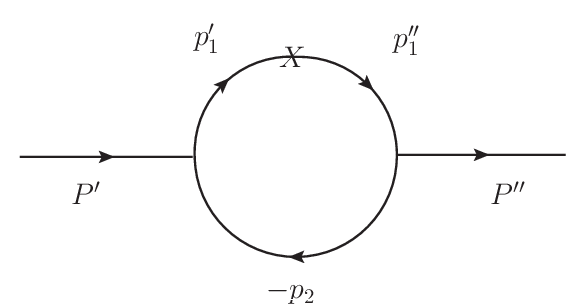}
 	\end{minipage}}
 	\caption{Feynman diagrams for $D^*_{(s)}$ decay (left) and transition
 		(right) amplitudes, where $P^{\prime(\prime\prime)}$ is the
 		incoming (outgoing) meson momentum, $p^{\prime(\prime\prime)}_1$
 		is the quark momentum, $p_2$ is the anti-quark momentum and X
 		denotes the vector or axial-vector transition vertex.}
 	\label{feyn}
 \end{figure}
Following the convention and calculation rules for the form factors of the transition $J/\Psi\to D$ in Ref. \cite{Sun2023}, one can express the decay amplitude in the lowest order for the transition $D^{*}_{(s)}\rightarrow P$, whose Fynman diagram is shown in Fig.\ref{feyn},
\be
\mathcal{B}_{\mu}^{D^{*}_{(s)} P}=-i^{3} \frac{N_{c}}{(2 \pi)^{4}} \int d^{4} p_{1}^{\prime} \frac{h_{D^{*}_{(s)}}^{\prime}\left(i h_{P}^{\prime \prime}\right)}{N_{1}^{\prime} N_{1}^{\prime \prime} N_{2}}
 S_{\mu \nu}^{D^{*}_{(s)} P} \varepsilon^{*\nu},
\en
where $N_{1}^{\prime(\prime \prime)}=p_{1}^{\prime(\prime \prime) 2}-m_{1}^{\prime (\prime\prime) 2}, N_{2}=p_{2}^{2}-m_{2}^{2} $ arise from the quark propagators. The $N_c$ represents the number of color degrees of freedom, which is conventionally set to 3. The trace $S_{\mu\nu}^{D^{*}_{(s)}P}$ can be obtained directly using Lorentz contraction,
\be
S_{\mu \nu}^{D^{*}_{(s)} P}&=&\left(S_{V}^{D^{*}_{(s)} P}-S_{A}^{D^{*}_{(s)} P}\right)_{\mu \nu}\non
&=&\operatorname{Tr}\left[\left(\gamma_{\nu}-\frac{1}{W_{V}^{\prime \prime}}\left(p_{1}^{\prime \prime}-p_{2}\right)_{\nu}\right)\left(p_{1}^{\prime \prime}
+m_{1}^{\prime \prime}\right)\left(\gamma_{\mu}-\gamma_{\mu} \gamma_{5}\right)\left(\not p_{1}^{\prime}+m_{1}^{\prime}\right) \gamma_{5}\left(-\not p_{2}
+m_{2}\right)\right].\;\;\;
\label{sptov}
\en
Its specific expression is listed in Appendix B. The covariant vertex function $h^{\prime}_{D^*_{(s)}}$ is defined as
\be
h_{D^*_{(s)}}^{\prime} &=&\left(M^{\prime 2}-M_{0}^{\prime 2}\right) \sqrt{\frac{x_{1} x_{2}}{N_{c}}} \frac{1}{\sqrt{2} \widetilde{M}_{0}^{\prime}} \varphi^{\prime},
\en
where $M^\prime$ refers to $m_{D^*_{(s)}}$,  $M'_0$ is the  kinetic invariant mass of the initial meson $D^*_{(s)}$ and can be expressed as the energies
 $e^{(\prime)}_i (i=1,2)$ of the constituent quark and anti-quark with masses (momentum fractions) being $m^\prime_1(x_1)$ and $m_2(x_2)$, respectively. Their definitions including
 the denominator $\widetilde{M}_{0}^{\prime}$ are given as follows
\be
M_{0}^{\prime 2} &=&\left(e_{1}^{\prime}+e_{2}\right)^{2}=\frac{p_{\perp}^{\prime 2}+m_{1}^{\prime 2}}{x_{1}}
+\frac{p_{\perp}^{\prime2}+m_{2}^{2}}{x_{2}}, \quad \widetilde{M}_{0}^{\prime}=\sqrt{M_{0}^{\prime 2}-\left(m_{1}^{\prime}-m_{2}\right)^{2}},\non
e_{i}^{(\prime)} &=&\sqrt{m_{i}^{(\prime) 2}+p_{\perp}^{\prime 2}+p_{z}^{\prime 2}}\;\;(i=1,2), \quad \quad p_{z}^{\prime}
=\frac{x_{2} M_{0}^{\prime}}{2}-\frac{m_{2}^{2}+p_{\perp}^{\prime 2}}{2 x_{2} M_{0}^{\prime}},
\en
The phenomenological Gaussian-type wave function $\varphi^{\prime}$ depicts the light-front momentum distribution amplitude for the S-wave mesons,
\be
\varphi^{\prime} &=&\varphi^{\prime}\left(x_{2}, p_{\perp}^{\prime}\right)=4\left(\frac{\pi}{\beta^{\prime 2}}\right)^{\frac{3}{4}}
\sqrt{\frac{d p_{z}^{\prime}}{d x_{2}}} \exp \left(-\frac{p_{z}^{\prime 2}+p_{\perp}^{\prime 2}}{2 \beta^{\prime 2}}\right),
\en
where $\beta^{\prime}$ is a phenomenological parameter and can be fixed by fitting the corresponding decay constant, and $p_{\perp}^{\prime}$ refers to the transverse momentum of the constituent quark. The expressions of the vertex functions $h^{\prime\prime}_P$ for our considered pseudoscalar mesons are similar. After expanding the trace $S_{\mu\nu}^{D^{*}_{(s)}P}$ using the Lortentz contraction, one can get the form factors $V^{D^{*}_{(s)}P}, A^{D^{*}_{(s)}P}_{0}, A^{D^{*}_{(s)}P}_{1}$ and $A^{D^{*}_{(s)}P}_{2}$ by matching to the coefficients given in Eq. (\ref{Dsp}). Their specific expressions are listed in Appendix B.

By combining the helicity amplitudes via the form factors, we can derive the differential widths of the semi-leptonic decays $D^{*}_{(s)}\to P\ell\nu_\ell$,
\begin{footnotesize}
\begin{eqnarray}
 \frac{d\Gamma_L}{dq^2}&=&(\frac{q^2-m_\ell^2}{q^2})^2\frac{ {\sqrt{\lambda(m_{D^{*}_{(s)}}^2,m_{P}^2,q^2)}} G_F^2 |V_{CKM}|^2} {384m_{D^{*}_{(s)}}^3\pi^3}
 \times \frac{1}{q^2} \Bigg\{ 3 m_\ell^2 \lambda(m_{D^{*}_{(s)}}^2,m_{P}^2,q^2) A_0^2(q^2)  \non
 &&+\frac{m_\ell^2+2q^2}{4m^2_{P}}  \left|
 (m_{D^{*}_{(s)}}^2-m_{P}^2-q^2)(m_{D^{*}_{(s)}}+m_{P})A_1(q^2)-\frac{\lambda(m_{D^{*}_{(s)}}^2,m_{P}^2,q^2)}{m_{D^{*}_{(s)}}+m_{P}}A_2(q^2)\right|^2
 \Bigg\},\label{eq:decaywidthlon}\;\;\;\;\;\;\\
\frac{d\Gamma_\pm}{dq^2}&=&(\frac{q^2-m_\ell^2}{q^2})^2\frac{ {\sqrt{\lambda(m_{D^{*}_{(s)}}^2,m_{P}^2,q^2)}} G_F^2 |V_{CKM}|^2} {384m_{D^{*}_{(s)}}^3\pi^3}
  \nonumber\\
 &&\;\;\times \Bigg\{ (m_\ell^2+2q^2) \lambda(m_{D^{*}_{(s)}}^2,m_{P}^2,q^2)\left|\frac{V(q^2)}{m_{D^{*}_{(s)}}+m_{P}}\mp
 \frac{(m_{D^{*}_{(s)}}+m_{P})A_1(q^2)}{\sqrt{\lambda(m_{D^{*}_{(s)}}^2,m_{P}^2,q^2)}}\right|^2
 \Bigg\},\label{eq:widthlon2}
\end{eqnarray}
\end{footnotesize}
where $\lambda(q^2)=\lambda(m^{2}_{D^{*}_{(s)}},m^{2}_{P},q^{2})=(m^{2}_{D^{*}_{(s)}}+m^{2}_{P}-q^{2})^{2}-4m^{2}_{D^{*}_{(s)}}m^{2}_{P}$, and $m_{\ell}$ is the mass of the lepton $\ell$. Although the electron and lepton masses are significantly small, we do not ignore them in the calculations to check the mass effects. The combined transverse and total differential decay widths are defined as
\be
\frac{d \Gamma_{T}}{d q^{2}}=\frac{d \Gamma_{+}}{d q^{2}}+\frac{d \Gamma_{-}}{d q^{2}}, \quad \frac{d \Gamma}{d q^{2}}=\frac{d \Gamma_{L}}{d q^{2}}+\frac{d \Gamma_{T}}{d q^{2}}.
\en

For the $D^{*}_{(s)}$ decays, it is meaningful to define the longitudinal polarization fraction owing to the existence of different polarizations
\be
f_{L}=\frac{\Gamma_{L}}{\Gamma_{L}+\Gamma_{+}+\Gamma_{-}}. \label{eq:fl}
\en
As to the forward-backward asymmetry, the analytical expression is defined as \cite{Sakaki},
\be
A_{FB} = \frac{\int^1_0 {d\Gamma \over dcos\theta} dcos\theta - \int^0_{-1} {d\Gamma \over dcos\theta} dcos\theta}
{\int^1_{-1} {d\Gamma \over dcos\theta} dcos\theta} = \frac{\int b_\theta(q^2) dq^2}{\Gamma_{D^{*}_{(s)}}},\label{eq:AFB}
\en
where $\theta$ is defined as the angle between the three-momenta of the lepton $\ell$ and the initial meson in the rest frame of $\ell\nu_{\ell}$. The function $b_{\theta}(q^2)$ refers to the angle coefficient and is expressed as \cite{Sakaki}
\be
b_\theta(q^2) &=& {G_F^2 |V_{CKM}|^2 \over 128\pi^3 m_{D^*_{(s)}}^3} q^2 \sqrt{\lambda(q^2)}
\left( 1 - {m_\ell^2 \over q^2} \right)^2 \left[ {1 \over 2}(H_{V,+}^2-H_{V,-}^2)+ {m_\ell^2 \over q^2} ( H_{V,0}H_{V,t} ) \right],
\label{eq:btheta2}
\en
where the helicity amplitudes for the $D^{*}_{(s)}\to P$ transitions are given as
\be
H_{V,\pm}\left(q^{2}\right)&=&\left(m_{D^*_{(s)}}+{m_{P}}\right) A_{1}\left(q^{2}\right) \mp \frac{\sqrt{\lambda\left(q^{2}\right)}}{m_{D^*_{(s)}}+m_{P}} V\left(q^{2}\right), \non
H_{V,0}\left(q^{2}\right)&=&\frac{m_{D^*_{(s)}}+m_{P}}{2 m_{D^*_{(s)}} \sqrt{q^{2}}}\left[-\left(m_{D^*_{(s)}}^{2}-m_{P}^{2}-q^{2}\right) A_{1}\left(q^{2}\right)+\frac{\lambda\left(q^{2}\right) A_{2}\left(q^{2}\right)}{\left(m_{D^*_{(s)}}+m_{P}\right)^{2}}\right],\non
H_{V,t}\left(q^{2}\right)&=&-\sqrt{\frac{\lambda\left(q^{2}\right)}{q^{2}}} A_{0}\left(q^{2}\right),
\en
where the subscript $V$ in each helicity amplitude refers to the $\gamma_\mu(1-\gamma_5)$ current.

Based on the effective Hamiltonian, the amplitudes for the decays $D^{*}_{(s)}\to PM_1$ with $M_1=\pi, K$ can be expressed as
\be
\mathcal{A}( D^{*}_{(s)}\to PM_1)=\left\langle P M_1\left|\mathcal{H}_{e f f}\right| D^{*}_{(s)}\right\rangle=\frac{G_F}{\sqrt{2}}V^*_{uq_1}V_{cq_2}a_i\left\langle M_1\left|J^{\mu}\right| 0\right\rangle\left\langle P \left|J_{\mu}\right|  D^{*}_{(s)}\right\rangle
\en
where $q_{1,2}=s, d$, the combination of the Wilson coefficients $a_{1}=C_1 +{C_2} /3$ and $a_{2}=C_2 +{C_2}/3$. As to the specific decay channels, the amplitudes are given as
\be
\mathcal{A}\left(D^{*+}_{s} \to \eta K^+ \right)&=&-\sqrt{2} G_{F} V_{us} V^{*}_{cs} a_{1} m_{D^*_s} \left(\epsilon \cdot p_{K}\right) f_{K} A^{D^*_s \eta_s}_{0} \sin\theta , \label{etak}\\
\mathcal{A}\left(D^{*+}_{s} \to \eta \pi^+ \right)&=&-\sqrt{2} G_{F} V_{ud} V^{*}_{cs} a_{1} m_{D^*_s} \left(\epsilon \cdot p_{\pi}\right) f_{\pi} A^{D^*_s \eta_s}_{0}\sin\theta,\\
\mathcal{A}\left(D^{*+} \to \eta K^+ \right)&=& \sqrt{2} G_{F} V_{us} V^{*}_{cd} a_{1} m_{D^*} \left(\epsilon \cdot p_{K}\right) f_{K} A^{D^* \eta_q}_{0}\cos\theta ,\\
\mathcal{A}\left(D^{*+} \to \eta \pi^+ \right)&=& \sqrt{2} G_{F} V_{ud} V^{*}_{cd} a_{1} m_{D^*} \left(\epsilon \cdot p_{\pi}\right) f_{\pi} A^{D^* \eta_q}_{0}\cos\theta ,\\
\mathcal{A}\left(D^{*0} \to \eta K^0 \right)&=& \sqrt{2} G_{F} V_{us} V^{*}_{cd} a_{1} m_{D^*} \left(\epsilon \cdot p_{K}\right) f_{K} A^{D^* \eta_q}_{0}\cos\theta , \label{etak0}\\
\mathcal{A}\left(D^{*+}_{s} \to K^0 K^+ \right)&=& \sqrt{2} G_{F} V_{us} V^{*}_{cd} a_{1} m_{D^*_s} \left(\epsilon \cdot p_{K}\right) f_{K} A^{D^*_s K}_{0}, \\
\mathcal{A}\left(D^{*+}_{s} \to K^0 \pi^+ \right)&=& \sqrt{2} G_{F} V_{ud} V^{*}_{cd} a_{1} m_{D^*_s} \left(\epsilon \cdot p_{\pi}\right) f_{\pi} A^{D^*_s K}_{0} ,\\
\mathcal{A}\left(D^{*+} \to \bar{K}^0 K^+ \right)&=& \sqrt{2} G_{F} V_{us} V^{*}_{cs} a_{1} m_{D^*} \left(\epsilon \cdot p_{K}\right) f_{K} A^{D^* K}_{0},\\
\mathcal{A}\left(D^{*+} \to \bar{K}^0 \pi^+ \right)&=& \sqrt{2} G_{F} V_{ud} V^{*}_{cs} m_{D^*} \left(\epsilon \cdot p_{\pi}\right)(a_{1}f_{\pi}A^{D^* K}_{0}+a_{2}f_{K}A^{D^* \pi}_{0}) ,\\
\mathcal{A}\left(D^{*0} \to \pi^- K^+ \right)&=& \sqrt{2} G_{F} V_{us} V^{*}_{cd} a_{1} m_{D^*} \left(\epsilon \cdot p_{K}\right) f_{K} A^{D^* \pi}_{0},\\
\mathcal{A}\left(D^{*0} \to \pi^- \pi^+ \right)&=& \sqrt{2} G_{F} V_{ud} V^{*}_{cd} a_{1} m_{D^*} \left(\epsilon \cdot p_{\pi}\right) f_{\pi} A^{D^* \pi}_{0},\\
\mathcal{A}\left(D^{*0} \to K^- K^+ \right)&=& \sqrt{2} G_{F} V_{us} V^{*}_{cs} a_{1} m_{D^*} \left(\epsilon \cdot p_{K}\right) f_{K} A^{D^* K}_{0},\\
\mathcal{A}\left(D^{*0} \to K^- \pi^+ \right)&=& \sqrt{2} G_{F} V_{ud} V^{*}_{cs} a_{1} m_{D^*} \left(\epsilon \cdot p_{\pi}\right) f_{\pi} A^{D^* K}_{0},
\en
where $\epsilon$ is the polarization four vector of the $D^{*}_{(s)}$ meson, and $\theta$ is the mixing angle between the two flavor states $\eta_s$ and $\eta_q$, which is defined as
\begin{equation}
	\left(
	\begin{array}{c}
		\eta \\
		\eta^{'}
	\end{array}
	\right)
	=
	\left(
	\begin{array}{cc}
		\cos\theta & -\sin\theta \\
		\sin\theta & \cos\theta
	\end{array}
	\right)
	\left(
	\begin{array}{c}
		\eta_q \\
		\eta_s
	\end{array}
	\right).
\end{equation}
where the mixing angle $\theta$ has been well determined as $\theta=39.3^{\circ}\pm 1.0^{\circ}$ \cite{Feldmann}. In Eqs. (\ref{etak}) - (\ref{etak0}), if one replaces $\eta$ with $\eta^\prime$ in the final states for each decay,  $-\sin\theta(\cos\theta)$ should be replaced with $\cos\theta(\sin\theta)$.

For the decays $D^{*}_{(s)}\to PV$ with $V$ being $\rho, K^*, \phi$, the hadronic matrix elements can be expressed as
\be
\mathcal{A}\left( D^{*}_{(s)}\to P V \right)=\left\langle P V\left|\mathcal{H}_{\mathrm{eff}}\right|  D^{*}_{(s)} \right\rangle=\frac{G_{F}}{\sqrt{2}} V_{c q_{1}}^{*} V_{u q_{2}} a_{1,2} H_{\lambda},
\en
where $\lambda=0, \mp$ denotes the helicity of vector meson, $G_{F}$ is the Fermi coupling constant, $V^{\ast}_{cq_1}V_{uq_2}$ is the product of the CKM matrix elements, and the helicity amplitudes $\mathcal{H}_{\lambda}=\left\langle V\left|J^{\mu}\right|0\right\rangle \left\langle P\left|J_{\mu}\right|D^{*}_{(s)}\right\rangle$ are given as follows
\be
H_{0} &\equiv&  \left\langle V\left(\varepsilon_{0}^{\prime}, p_{V}\right)\left|\bar{q}_1 \gamma^{\mu} u\right| 0\right\rangle\left\langle P\left(p_{P}\right)\left|\bar{c} \gamma_{\mu}\left(1-\gamma_{5}\right) q_2\right|  D^{*}_{(s)}\left(\varepsilon_{0}, p_{ D^{*}_{(s)}}\right)\right\rangle \non&&
=\frac{i f_{V}}{2 m_{ D^{*}_{(s)}}} \Bigg[\left(m_{ D^{*}_{(s)}}^{2}-m_{P}^{2}+m_{V}^{2}\right)\left(m_{ D^{*}_{(s)}}+m_{P}\right) A_{1}^{ D^{*}_{(s)} P}\left(m_{V}^{2}\right) \non&&
 +\frac{4 m_{ D^{*}_{(s)}}^{2} p_{c}^{2}}{m_{ D^{*}_{(s)}}+m_{P}} A_{2}^{ D^{*}_{(s)}  P}\left(m_{V}^{2}\right)\Bigg], \\
H_{\mp} &\equiv& \left\langle V\left(\varepsilon_{\mp}^{\prime}, p_{V}\right)\left|\bar{q}_1 \gamma^{\mu} u\right| 0\right\rangle\left\langle P\left(p_{P}\right)\left|\bar{c} \gamma_{\mu}\left(1-\gamma_{5}\right) q_2\right|  D^{*}_{(s)}\left(\varepsilon_{\mp}, p_{ D^{*}_{(s)}}\right)\right\rangle \non&&
=i f_{V} m_{V}\Bigg[-\left(m_{D^{*}_{(s)}}+m_{P}\right) A_{1}^{D^{*}_{(s)}P}\left(m_{V}^{2}\right) \mp \frac{2 m_{D^{*}_{(s)}} p_{c}}{m_{D^{*}_{(s)}}+m_{P}} V^{D^{*}_{(s)}P}  \left(m_{V}^{2}\right)\Bigg].
\en
\section{Numerical results and discussions} \label{numer}
\subsection{Transition Form Factors}

\begin{table}[H]
\caption{Values of the input parameters \cite{ChangYang,ChengHwang,Zyla,Sun2023,EbertD,WangShen,YuLi}}.
\label{table1}
\begin{tabular*}{16.5cm}{@{\extracolsep{\fill}}l|ccccc}
  \hline\hline
\textbf{Masses(\text{GeV})}
&$m_{c}=1.4$&$m_{s}=0.37$&$m_{u,d}=0.25$&$m_e=0.000511$    \\[1ex]
&$m_{\mu}=0.106$&$m_{\tau}=1.777$&$m_{\pi}=0.140$&$m_{\rho}=0.770$\\[1ex]
&$m_{K}=0.494$&$m_{K^{0}}=0.498$&$m_{\eta}=0.548$& $ m_{\eta^{'}}=0.958$\\[1ex]
&$m_{\phi}=1.019$&$m_{D^{*0}}=2.007$&$m_{D^{*+}}=2.01$&$m_{D_{s}^{*}}=2.112$\\[1ex]
\hline
\end{tabular*}
\begin{tabular*}{16.5cm}{@{\extracolsep{\fill}}l|ccccc}
  \hline
\multirow{2}{*}{\textbf{CKM}}&$V_{cd}=0.221\pm0.004$&$V_{us}=0.2243\pm0.0008$\\[1ex]
$$&$V_{cs}=0.975\pm0.006$& $V_{ud}=0.97373\pm0.00031$ \\[1ex]
\hline
\end{tabular*}
\begin{tabular*}{16.5cm}{@{\extracolsep{\fill}}l|cccccc}
\hline
 &$f_{\pi}=0.132$ & $f_{K}=0.16$&$f_{D^{*}}=0.310^{+0.046}_{-0.046}$  \\[1ex]
\textbf{ Decay constants(\text{GeV})} &$f_{\eta_{q}}=0.141$  &$f_{\eta_{s}}=0.177$  & $f_{D_{s}^{*}}=0.301^{+0.045}_{-0.045}$\\[1ex]
&$f_{K^{*}}=0.217$ &$f_{\rho}=0.209$ &$f_{\phi}=0.229$ \\[1ex]
\hline\hline
\end{tabular*}
\begin{tabular*}{16.5cm}{@{\extracolsep{\fill}}l|ccccc}
\textbf{Shape parameters(\text{GeV})}&$\beta_{D^{*}}=0.474^{+0.042}_{-0.046}$&$\beta_{D_{s}^{*}}=0.466^{+0.042}_{-0.046}$&$\beta_{K}=0.394^{+0.003}_{-0.003}$\\[1ex]
&$\beta_{\eta_{q}}=0.374^{+0.02}_{-0.03}$&$\beta_{\eta_{s}}=0.404^{+0.01}_{-0.02}$&$\beta_{\pi}=0.328^{+0.002}_{-0.004}$\\[1ex]
\hline\hline
\end{tabular*}
\begin{tabular*}{16.5cm}{@{\extracolsep{\fill}}l|ccc}
\textbf{Full widths}&$\Gamma_{D^{*0}}=(55.9^{+5.9}_{-5.4})\text{keV}$&$\Gamma_{D^{*+}}=(83.4\pm1.8)\text{keV}$&$\Gamma_{D_{s}^{*+}}=(121.9^{+69.6}_{-52.2})\text{eV}$\\[1ex]
\hline\hline
\end{tabular*}
\end{table}

The input parameters, such as the constituent quark masses, the masses of the initial and final mesons, the Cabibbo-Kobayashi-Maskawa (CKM) matrix elements, the shape parameters fitted by the decay constants, and $D^{*}_{(s)}$ meson lives, are listed in Table \ref{table1}.
Based on the input parameters given in Table \ref{table1}, one can obtain the numerical results of the transition form factors at $q^2 = 0$ shown in Tables \ref{table2}. The uncertainties originate from the shape parameters of the initial and final state mesons.

All the calculations are performed within the $q^+ = 0$ reference frame, where the form factors can only be obtained at space-like momentum transfers $q^2 = - q^2_\bot \leq 0$. Those parameterized form factors are extrapolated from the space-like region to the time-like region by using following expression,
\be
F\left(q^{2}\right)=\frac{F(0)}{1-a q^{2} / m^{2}+b q^{4} / m^{4}},
\en
where $F(q^{2})$ denotes different form factors $V(q^2), A_0(q^2), A_1(q^2)$ and $A_2(q^2)$, and $m$ represents the initial meson mass. The values of $a$ and $b$ can be obtained by performing a 3-parameter
fit to the form factors in the range -15GeV$^{2} \leq q^{2} \leq 0$. $F(q^{2}_{max})$ is defined as the value of the form factor evaluated at the maximum value of $q^2$. In Tables \ref{table2}, we list the computed values of $F(q^{2}_{max})$, $a$ and $b$ for various decay channels.

\begin{table}[H]
\caption{Form factors of the transitions $D^*\to K,\pi, \eta_{q}$ and $D^*_s\to K, \eta_{s}$ at $q^{2}=0$. Numerical results of $F(q^{2}_{max})$, $a$ and $b$ obtained from the calculations. The uncertainties originate from the shape parameters of the initial and final state mesons.}
\begin{center}
\scalebox{1.15}{
\begin{tabular}{|c|c|ccccc|}
\hline\hline
Transitions&Ref.&$$&$V$&$A_{0}$&$A_{1}$&$A_{2}$\\
\hline\hline
$D^{*}\rightarrow K$&This work&$F(0)$&$0.96^{+0.01}_{-0.02}$&$0.64^{+0.01}_{-0.02}$&$0.78^{+0.01}_{-0.02}$&$0.40^{+0.01}_{-0.02}$\\
\hline
$$&$$&$F(q^{2}_{max})$&$0.98^{+0.02}_{-0.01}$&$0.75^{+0.03}_{-0.00}$&$0.88^{+0.02}_{-0.01}$&$0.45^{+0.02}_{-0.02}$\\
$$&$$&$a$&$0.35^{+0.02}_{-0.03}$&$0.26^{+0.01}_{-0.01}$&$0.22^{+0.00}_{-0.01}$&$0.30^{+0.02}_{-0.02}$\\
$$&$$&$b$&$0.57^{+0.06}_{-0.05}$&$0.04^{+0.01}_{-0.01}$&$0.03^{+0.01}_{-0.01}$&$0.22^{+0.01}_{-0.01}$\\
\hline\hline
$D^{*}\rightarrow \pi$&This work&$F(0)$&$0.77^{+0.02}_{-0.02}$&$0.57^{+0.01}_{-0.01}$&$0.75^{+0.00}_{-0.01}$&$0.37^{+0.01}_{-0.01}$\\
\hline
$$&$$&$F(q^{2}_{max})$&$0.57^{+0.02}_{-0.02}$&$0.74^{+0.02}_{-0.01}$&$0.92^{+0.01}_{-0.00}$&$0.37^{+0.01}_{-0.01}$\\
$$&$$&$a$&$0.29^{+0.04}_{-0.05}$&$0.31^{+0.03}_{-0.01}$&$0.25^{+0.02}_{-0.01}$&$0.31^{+0.06}_{-0.03}$\\
$$&$$&$b$&$0.80^{+0.11}_{-0.09}$&$0.07^{+0.01}_{-0.01}$&$0.04^{+0.01}_{-0.01}$&$0.34^{+0.02}_{-0.01}$\\
\hline\hline
$D^{*}_s\rightarrow K$&This work&F(0)&$0.98^{+0.01}_{-0.01}$&$0.53^{+0.02}_{-0.03}$&$0.66^{+0.02}_{-0.03}$&$0.31^{+0.02}_{-0.03}$\\
\hline
$$&$$&$F(q^{2}_{max})$&$0.80^{+0.01}_{-0.00}$&$0.60^{+0.03}_{-0.02}$&$0.74^{+0.03}_{-0.02}$&$0.30^{+0.03}_{-0.02}$\\
$$&$$&$a$&$0.25^{+0.04}_{-0.05}$&$0.26^{+0.01}_{-0.02}$&$0.24^{+0.01}_{-0.01}$&$0.23^{+0.04}_{-0.05}$\\
$$&$$&$b$&$1.06^{+0.13}_{-0.11}$&$0.11^{+0.02}_{-0.02}$&$0.08^{+0.02}_{-0.01}$&$0.35^{-0.02}_{-0.02}$\\
\hline\hline
$D^{*}\rightarrow \eta_{q} $&This work&$F(0)$&$0.96^{+0.02}_{-0.02}$&$0.56^{+0.01}_{-0.02}$&$0.67^{+0.01}_{-0.02}$&$0.36^{+0.01}_{-0.02}$\\
\hline
$$&$$&$F(q^{2}_{max})$&$0.95^{+0.02}_{-0.02}$&$0.64^{+0.02}_{+0.01}$&$0.75^{+0.02}_{-0.01}$&$0.39^{+0.02}_{-0.02}$\\
$$&$$&$a$&$0.33^{+0.02}_{-0.03}$&$0.25^{+0.01}_{-0.01}$&$0.21^{+0.00}_{-0.01}$&$0.27^{+0.02}_{-0.03}$\\
$$&$$&$b$&$0.66^{+0.07}_{-0.06}$&$0.05^{+0.01}_{-0.01}$&$0.03^{+0.01}_{-0.01}$&$0.23^{+0.01}_{-0.01}$\\
\hline\hline
$D^{*}_s\rightarrow \eta_s $&This work&$F(0)$&$1.17^{+0.00}_{-0.02}$&$0.63^{+0.01}_{-0.02}$&$0.70^{+0.01}_{-0.02}$&$0.45^{+0.02}_{-0.03}$\\
\hline
$$&$$&$F(q^{2}_{max})$&$1.17^{+0.02}_{-0.00}$&$0.68^{+0.03}_{-0.02}$&$0.75^{+0.02}_{-0.01}$&$0.47^{+0.03}_{-0.03}$\\
$$&$$&$a$&$0.28^{+0.04}_{-0.05}$&$0.31^{+0.01}_{-0.01}$&$0.29^{+0.00}_{-0.01}$&$0.29^{+0.03}_{-0.03}$\\
$$&$$&$b$&$0.95^{+0.12}_{-0.10}$&$0.11^{+0.02}_{-0.02}$&$0.09^{+0.02}_{-0.02}$&$0.38^{+0.03}_{-0.02}$\\
\hline\hline
\end{tabular}}\label{table2}
\end{center}
\end{table}

\begin{table}[H]
\caption{Form factors of the transitions $D^*\to K,\pi$ at $q^{2}=0$, together with other theoretical results. The uncertainties originate from the shape parameters of the initial and final state mesons.}
\begin{center}
\scalebox{1.15}{
\begin{tabular}{|c|c|ccccc|}
\hline\hline
Transitions&Ref.&$$&$V$&$A_{0}$&$A_{1}$&$A_{2}$\\
\hline\hline
$D^{*}\rightarrow K$&This work&$F(0)$&$0.96^{+0.01}_{-0.02}$&$0.64^{+0.01}_{-0.02}$&$0.78^{+0.01}_{-0.02}$&$0.40^{+0.01}_{-0.02}$\\
\hline
$$&Ref.\cite{ChangYang}&$$&$1.04$&$0.78$&$0.85$&$0.68$\\
\hline\hline
$D^{*}\rightarrow \pi$&This work&$F(0)$&$0.77^{+0.02}_{-0.02}$&$0.57^{+0.01}_{-0.01}$&$0.75^{+0.00}_{-0.01}$&$0.37^{+0.01}_{-0.01}$\\
\hline
$$&Ref.\cite{ChangYang}&$$&$0.92$&$0.68$&$0.74$&$0.61$\\
\hline\hline
\end{tabular}}\label{table2.2}
\end{center}
\end{table}

\begin{figure}[H]
\vspace{0.60cm}
  \centering
  \subfigure[]{\includegraphics[width=0.28\textwidth]{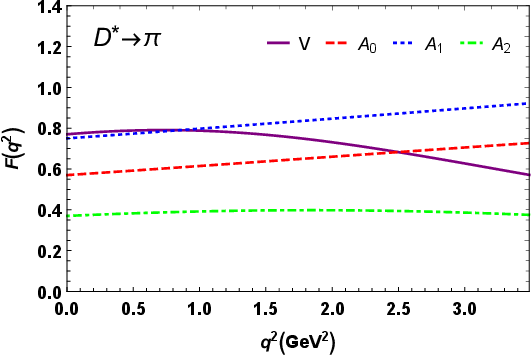}\quad}
  \subfigure[]{\includegraphics[width=0.28\textwidth]{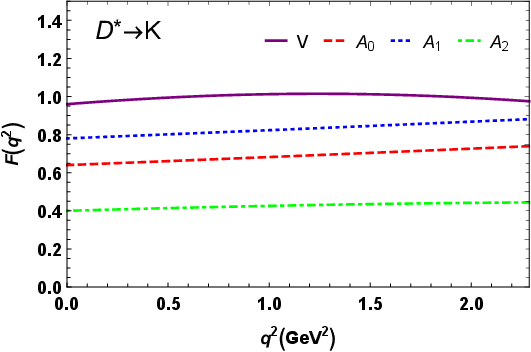}\quad}
   \subfigure[]{\includegraphics[width=0.28\textwidth]{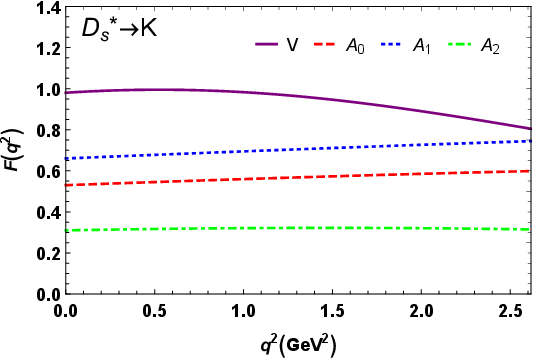}\quad}
  \subfigure[]{\includegraphics[width=0.28\textwidth]{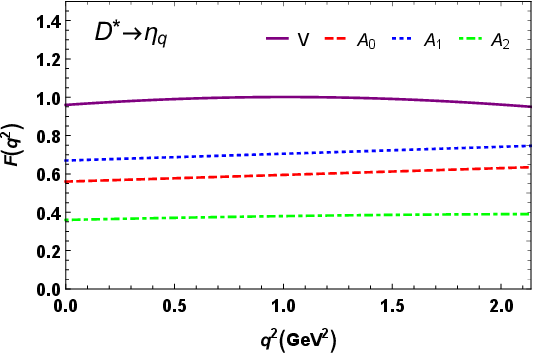}\quad}
  \subfigure[]{\includegraphics[width=0.28\textwidth]{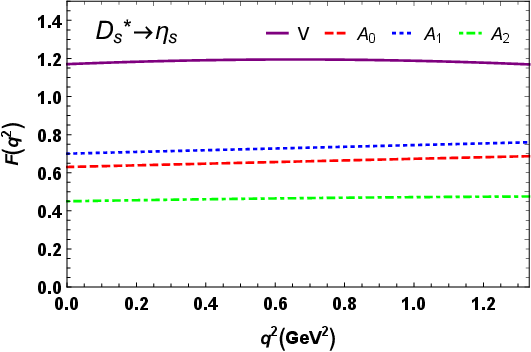}\quad}
\caption{(color online) Form factors $V(q^2)$, $A_{0}(q^2)$, $A_{1}(q^2)$, $A_{2}(q^2)$ of the transitions $D^{*}_{(s)} \rightarrow \pi, K, \eta_q, \eta_s$ .}\label{fig2}
\end{figure}

In Table \ref{table2.2}, we compare the numerical values of the form factors at maximum recoil $(q^{2} = 0)$ with those obtained in Ref. \cite{ChangYang}. Our predictions for the form factors of the transitions $D^{*}\rightarrow K, \pi$ are comparable to the previous LFQM calculations \cite{ChangYang} within errors. The difference between these two works is partially caused by the input parameters.
We plot the $q^{2}$-dependence of the form factors of the transitions $D^{*}_{(s)} \rightarrow P$ in Fig. \ref{fig2}.

Note that the $V(q^2)$ curves for the transitions $D^*\to K, \eta_q$ and $D^*_s \to \eta_s$ exhibit nearly flat in the whole $q^2$ region. The larger phase space for the transitions $D^*\to\pi$ and $D^*_s\to K$ renders the form factors more sensitive to variations in $q^2$.

\subsection{Semi-leptonic decays}

Based the form factors and the helicity amplitudes provided in the previous section, the branching ratios of the semi-leptonic decays $D^{*}_{(s)}\to P\ell\nu_{\ell}$ can be obtained and are listed in Table \ref{table4}, where the first error arises from the decay widths of $D^*_{(s)}$, and the second uncertainty originates from the shape parameters of the initial and final mesons.

\begin{table}[H]
	\caption{Branching  ratios of the semi-leptonic decays $D^{*}_{(s)}\to P\ell\nu_{\ell}$ .}
	\begin{center}
		\scalebox{0.8}{
			\begin{tabular}{|c|c|c|c|c|}
				\hline\hline
				$$&$10^{-9}\times \mathcal{B}r(D^{*0}\rightarrow \pi ^{-}e^{+}\nu_{e})$&$10^{-9}\times \mathcal{B}r(D^{*0}\rightarrow \pi^{-} \mu^{+}\nu_{\mu})$&$10^{-9}\times \mathcal{B}r(D^{*0}\rightarrow K^{-}e^{+}\nu_{e})$&$10^{-9}\times \mathcal{B}r(D^{*0}\rightarrow K^{-}\mu^{+}\nu_{\mu})$\\
				\hline
				This work&$2.79^{+0.30+0.21}_{-0.27-0.03}$&$2.64^{+0.28+0.19}_{-0.25-0.04}$&$7.93^{+0.84+0.07}_{-0.76-0.36}$&$7.53^{+0.81+0.07}_{-0.72-0.34}$\\
				\hline
				$$&$10^{-9}\times \mathcal{B}r(D^{*+}\rightarrow \pi ^{0}e^{+}\nu_{e})$&$10^{-10}\times \mathcal{B}r(D^{*+}\rightarrow \pi^{0} \mu^{+}\nu_{\mu})$&$10^{-14}\times \mathcal{B}r(D^{*+}\rightarrow \pi^{0}\tau^{+}\nu_{\tau})$&$10^{-14}\times \mathcal{B}r(D^{*0}\rightarrow \pi^{-}\tau^{+}\nu_{\tau})$\\
				\hline
				This work&$1.00^{+0.02+0.07}_{-0.02-0.01}$&$9.50^{+0.21+0.69}_{-0.20-0.12}$&$2.25^{+0.05+0.06}_{-0.05-0.04}$&$4.96^{+0.53+0.14}_{-0.47-0.08}$\\
				\hline
				$$&$10^{-9}\times \mathcal{B}r(D^{*+}\rightarrow \bar{K}^0e^{+}\nu_{e})$&$10^{-9}\times \mathcal{B}r(D^{*+}\rightarrow \bar{K}^0\mu^{+}\nu_{\mu})$&$10^{-7}\times \mathcal{B}r(D_{s}^{*+}\rightarrow K^{0}e^{+}\nu_{e})$&$10^{-7}\times \mathcal{B}r(D_{s}^{*+}\rightarrow K^{0}\mu^{+}\nu_{\mu})$\\
				\hline
				This work&$5.29^{+0.12+0.05}_{-0.11-0.24}$&$5.03^{+0.11+0.05}_{-0.11-0.23}$&$1.97^{+0.21+0.06}_{-0.16-0.09}$&$1.87^{+0.20+0.06}_{-0.16-0.09}$\\
				\hline
				$$&$10^{-11}\times \mathcal{B}r(D^{*+}\rightarrow \eta e^{+}\nu_{e})$&$10^{-11}\times \mathcal{B}r(D^{*+}\rightarrow \eta \mu^{+}\nu_{\mu})$&$10^{-12}\times \mathcal{B}r(D^{*+}\rightarrow \eta ^{\prime}e^{+}\nu_{e})$&$10^{-12}\times \mathcal{B}r(D^{*+}\rightarrow \eta^{\prime}\mu^{+}\nu_{\mu})$\\
				\hline
				This work&$5.03^{+0.11+0.18}_{-0.11-0.08}$&$4.79^{+0.11+0.17}_{-0.10-0.08}$&$7.58^{+0.17+0.20}_{-0.16-0.16}$&$7.06^{+0.16+0.27}_{-0.15-0.16}$\\
				\hline
				$$&$10^{-6}\times \mathcal{B}r(D_{s}^{*+}\rightarrow \eta e^{+}\nu_{e})$&$10^{-6}\times \mathcal{B}r(D_{s}^{*+}\rightarrow \eta \mu^{+}\nu_{\mu})$&$10^{-7}\times \mathcal{B}r(D_{s}^{*+}\rightarrow \eta ^{\prime}e^{+}\nu_{e})$&$10^{-7}\times \mathcal{B}r(D_{s}^{*+}\rightarrow \eta^{\prime}\mu^{+}\nu_{\mu})$\\
				\hline
				This work&$1.46^{+0.16+0.03}_{-0.12-0.09}$&$1.41^{+0.15+0.02}_{-0.12-0.09}$&$5.08^{+0.55+0.10}_{-0.42-0.37}$&$4.80^{+0.52+0.10}_{-0.40-0.36}$\\
				\hline\hline
		\end{tabular}}\label{table4}
	\end{center}
\end{table}

The branching ratios of the semi-leptonic $D^{*+}_s\to P \ell^+\nu_\ell$ decays are in the order of $10^{-7}\sim 10^{-6}$, which are much larger than those of the semi-leptonic $D^{*0(+)}\to P^{-(0)} \ell^+\nu_\ell$ decays within the range $10^{-12}\sim 10^{-9}$. This is mainly because the decay width of the meson $D^*_s$ is much smaller than that of the meson $D^*$. In these semi-leptonic $D^*_s$ decays, the decays $D^{*+}_s\to \eta \ell^{+}\nu_{\ell}$ have the largest
branching ratios, which can be detected by the present experiments, whereas the branching ratios of the decays $D^{*+}_s\to \eta^\prime \ell^{+}\nu_{\ell}$ are approximately three times smaller. This is mainly because of the smaller phase space. Although two other semi-leptonic $D^*_s$ decays $D^{*+}_s\to K^0 \ell^{+}\nu_{\ell}$ induced by the $c\to d$ transition have a much smaller CKM matrix element $V_{cd}$ than $V_{cs}$, the large phase space can reduce the gap of the branching ratios between the decays $D^{*+}_s\to K^0 \ell^{+}\nu_{\ell}$ and $D^{*+}_s\to \eta^\prime \ell^{+}\nu_{\ell}$. As the $\tau$ lepton mass is much greater than the $e$ and $\mu$ lepton masses,  $\tau$ can only be produced through the decays $D^{*0(+)}\rightarrow \pi^{-(0)}\tau^{+}\nu_{\tau}$ with tiny branching ratios, which are only within the order of $10^{-14}$.
In other words, these considered semi-leptonic $D^*_s$ decays have large branching ratios, which are larger than the order of$10^{-7}$ and can even reach up to the order of $10^{-6}$. Hence, they are most likely to be detected by future high-luminosity experiments, such as the STCF, BESIII and LHCb. Furthermore,
it is meaningful to define the ratios of the branching fractions, which can be estimated through the total widths of the initial mesons and the CKM matrix elements, for example,
\be
\frac{\mathcal{B}r(D_{s}^{*+}\rightarrow \eta e^{+}\nu_{e})}{\mathcal{B}r(D^{*+}\rightarrow \eta e^{+}\nu_{e})}&\approx & \frac{2V_{cs}^2 \Gamma_{D^{*+}}}{V_{cd}^2 \Gamma_{D_s^{*+}}} \approx 2.7 \times 10^4, \\
\frac{\mathcal{B}r(D_{s}^{*+}\rightarrow K^0 e^{+}\nu_{e})}{\mathcal{B}r(D^{*+}\rightarrow \bar{K}^0 e^{+}\nu_{e})}&\approx &\frac{V_{cd}^2 \Gamma_{D^{*+}}}{V_{cs}^2 \Gamma_{D_s^{*+}}} \approx 35.
\en
These ratios are consistent with the results calculated from the branching ratios given in Table \ref{table4}, which can be tested in the future experiments.
\subsection{Physical observables for semi-leptonic decays}
\begin{figure}[H]
	\vspace{0.6cm}
	\centering
	\subfigure[]{\includegraphics[width=0.35\textwidth]{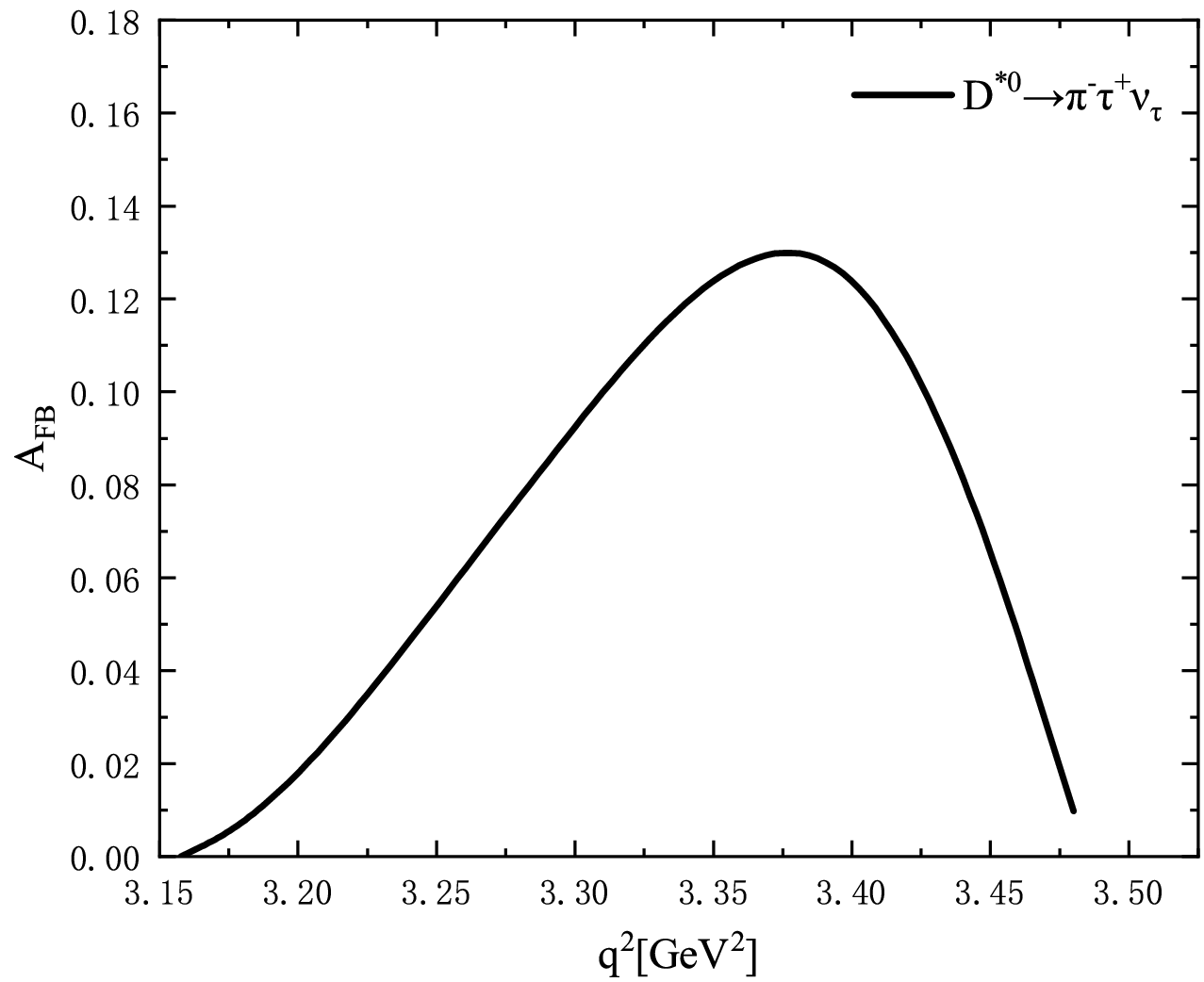}\quad}
	\subfigure[]{\includegraphics[width=0.35\textwidth]{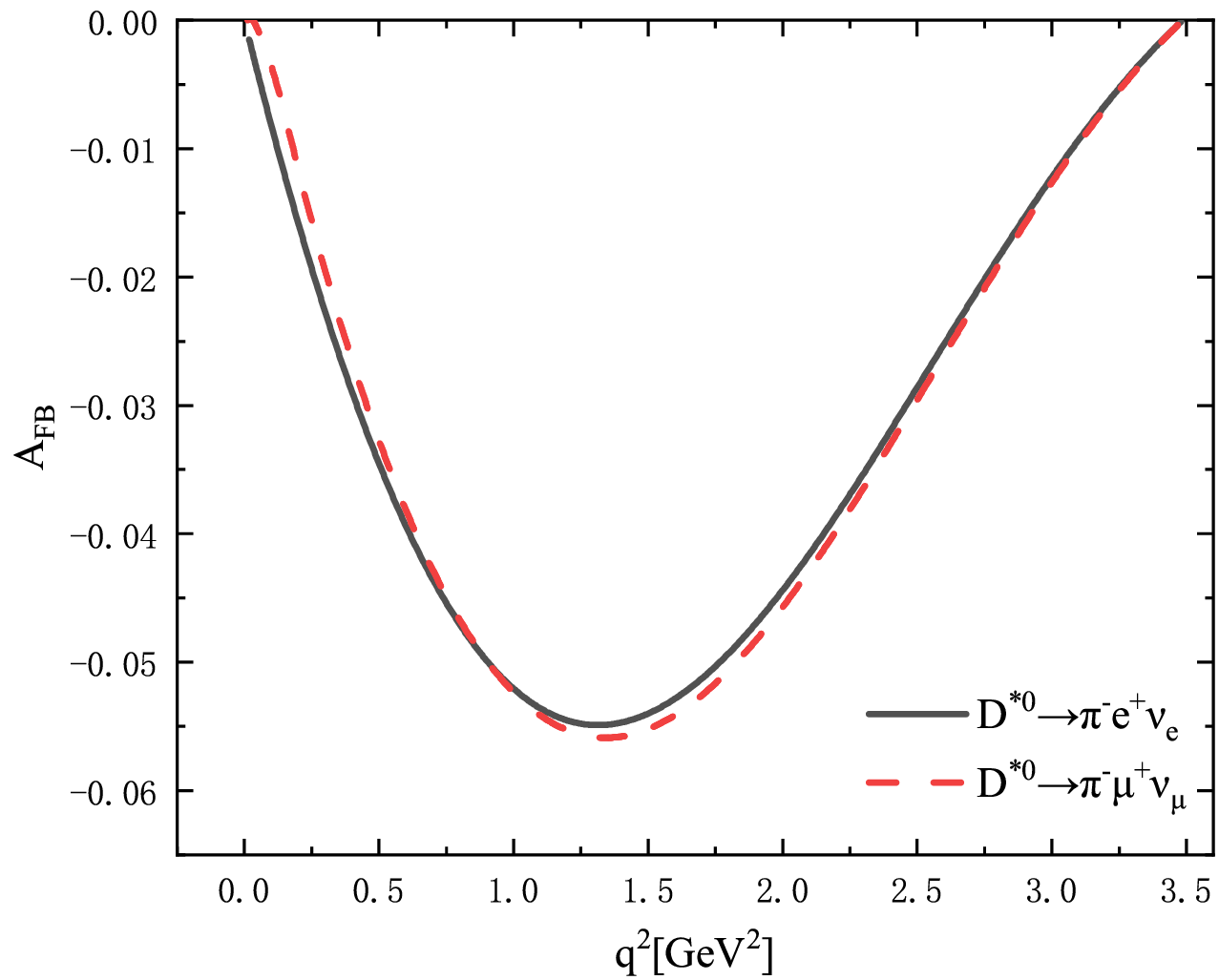}\quad}\\
	\subfigure[]{\includegraphics[width=0.35\textwidth]{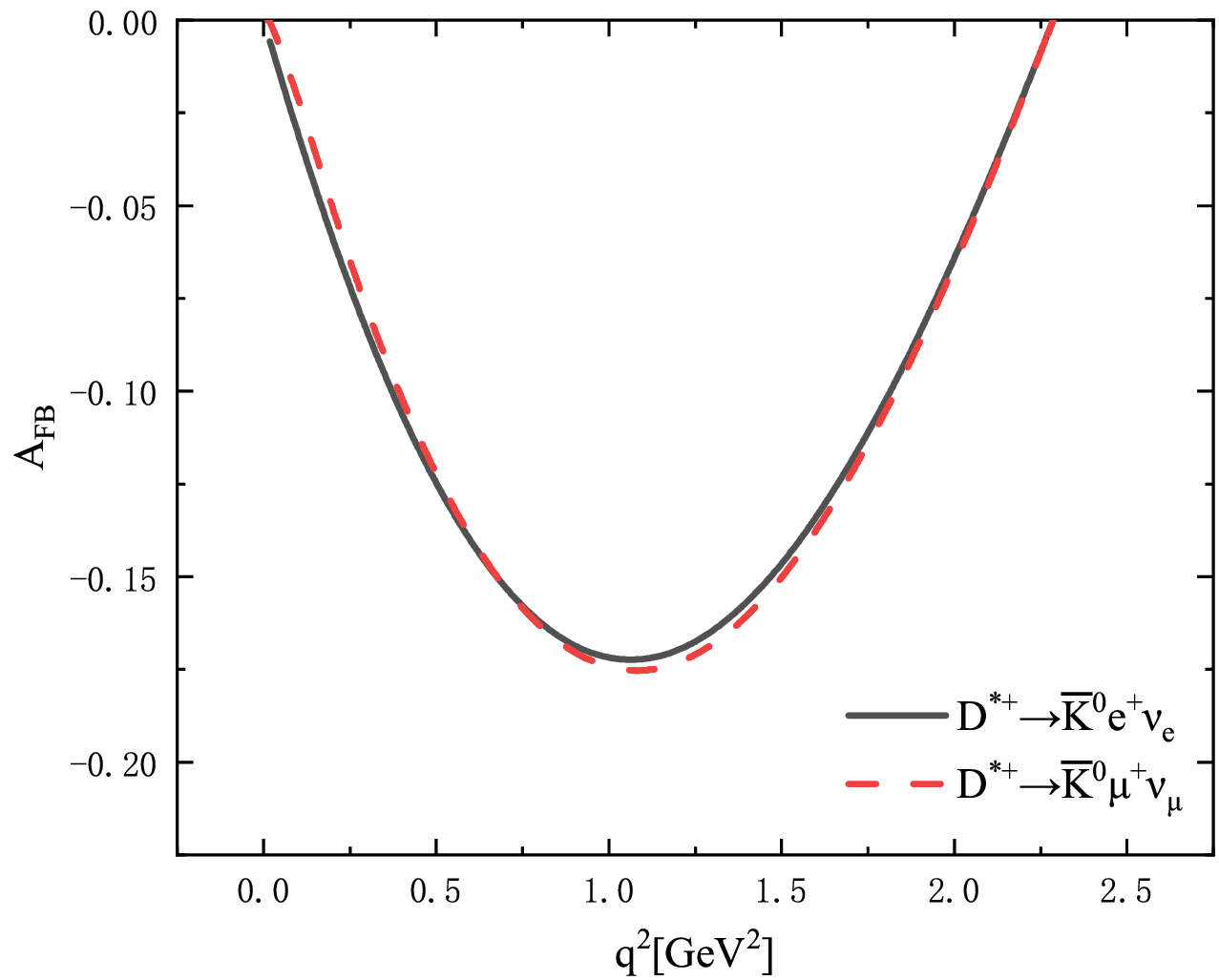}\quad}\;\;
	\subfigure[]{\includegraphics[width=0.35\textwidth]{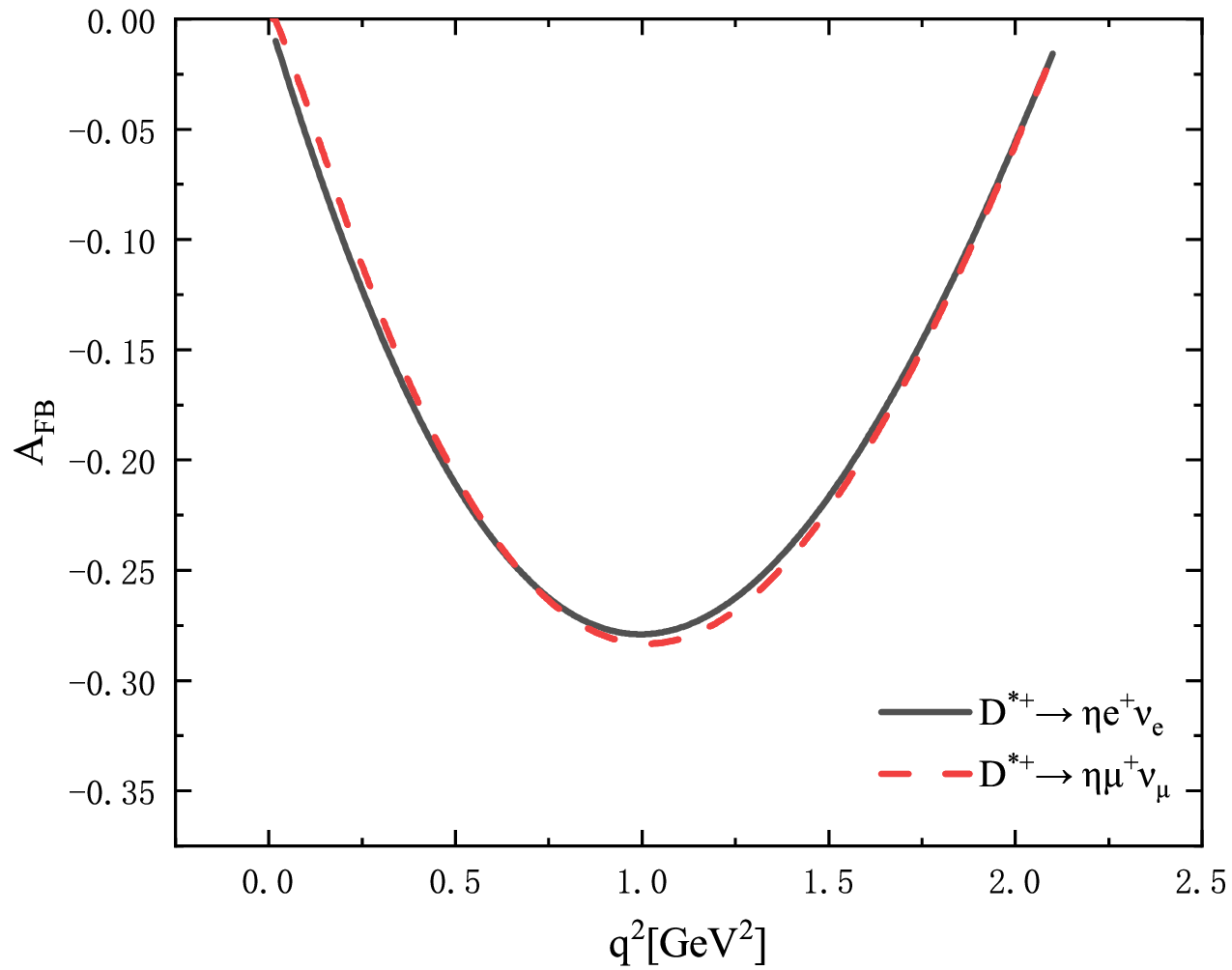}\quad}
	\caption{ (color online) $q^{2}$-dependences of the forward-backward asymmetries $A_{FB}$ for the decays $D^{*0}\to \pi^-\tau^+\nu_{\tau}$ (a), $D^{*0}\to \pi^-\ell^+\nu_{\ell}$ (b), $D^{*+}\to \bar K^0\ell^+\nu_{\ell}$ (c) and $D^{*+}\to \eta\ell^+\nu_{\ell}$ (d).}\label{AFB.}
\end{figure}

\begin{figure}[H]
	\vspace{0.60cm}
	\centering
	\subfigure[]{\includegraphics[width=0.22\textwidth]{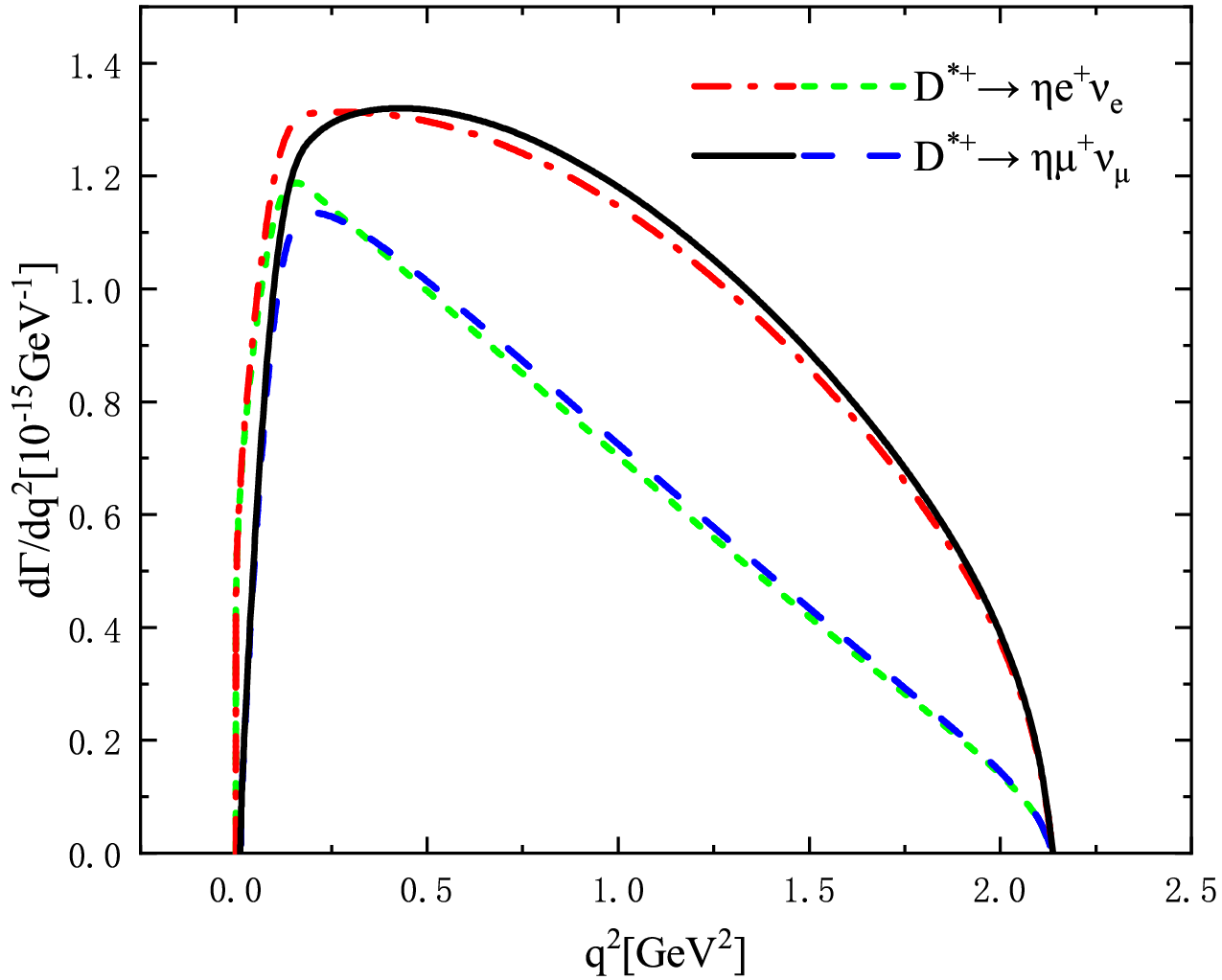}\quad}
	\subfigure[]{\includegraphics[width=0.22\textwidth]{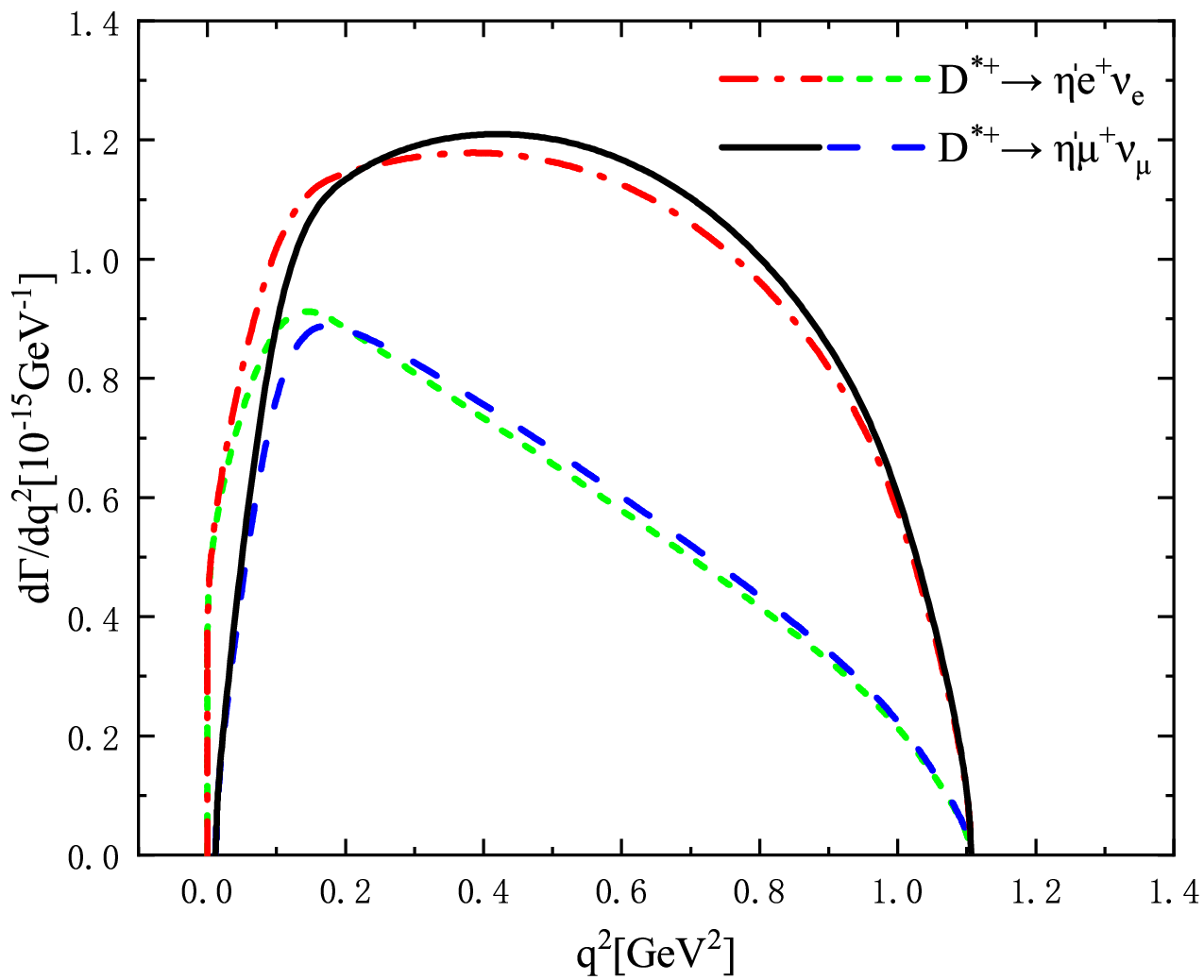}\quad}
	\subfigure[]{\includegraphics[width=0.22\textwidth]{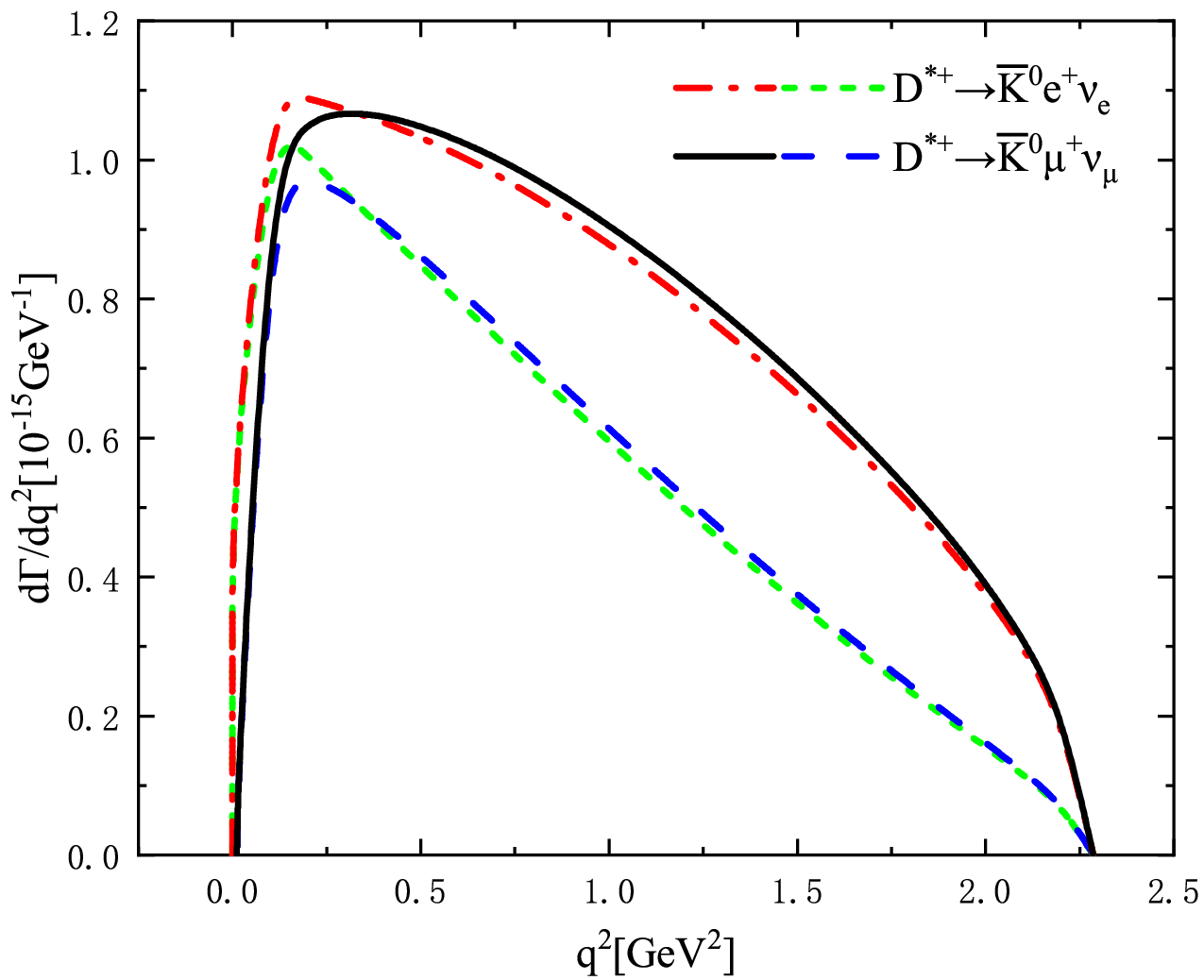}\quad}
	\subfigure[]{\includegraphics[width=0.22\textwidth]{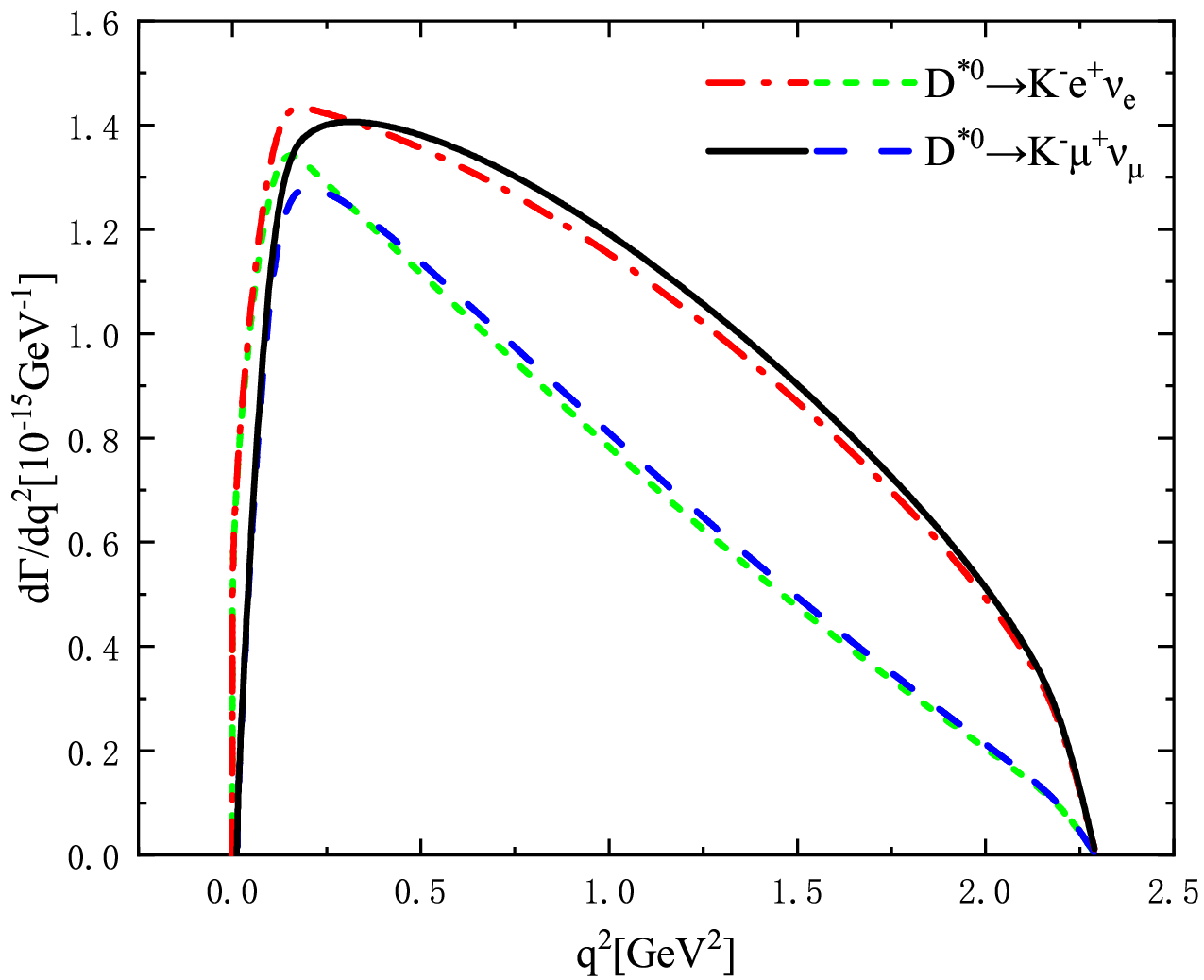}\quad}
	\subfigure[]{\includegraphics[width=0.22\textwidth]{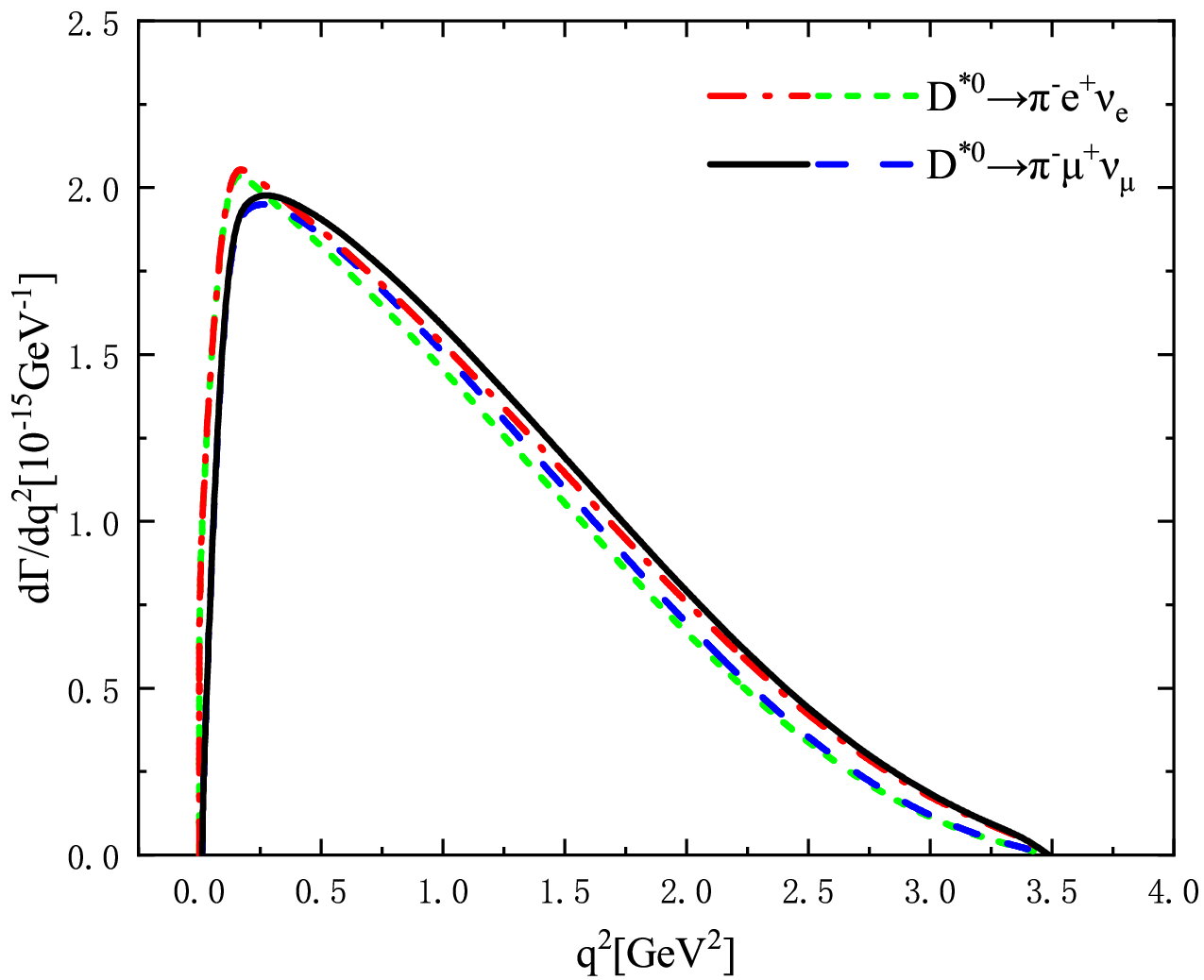}\quad}
	\subfigure[]{\includegraphics[width=0.22\textwidth]{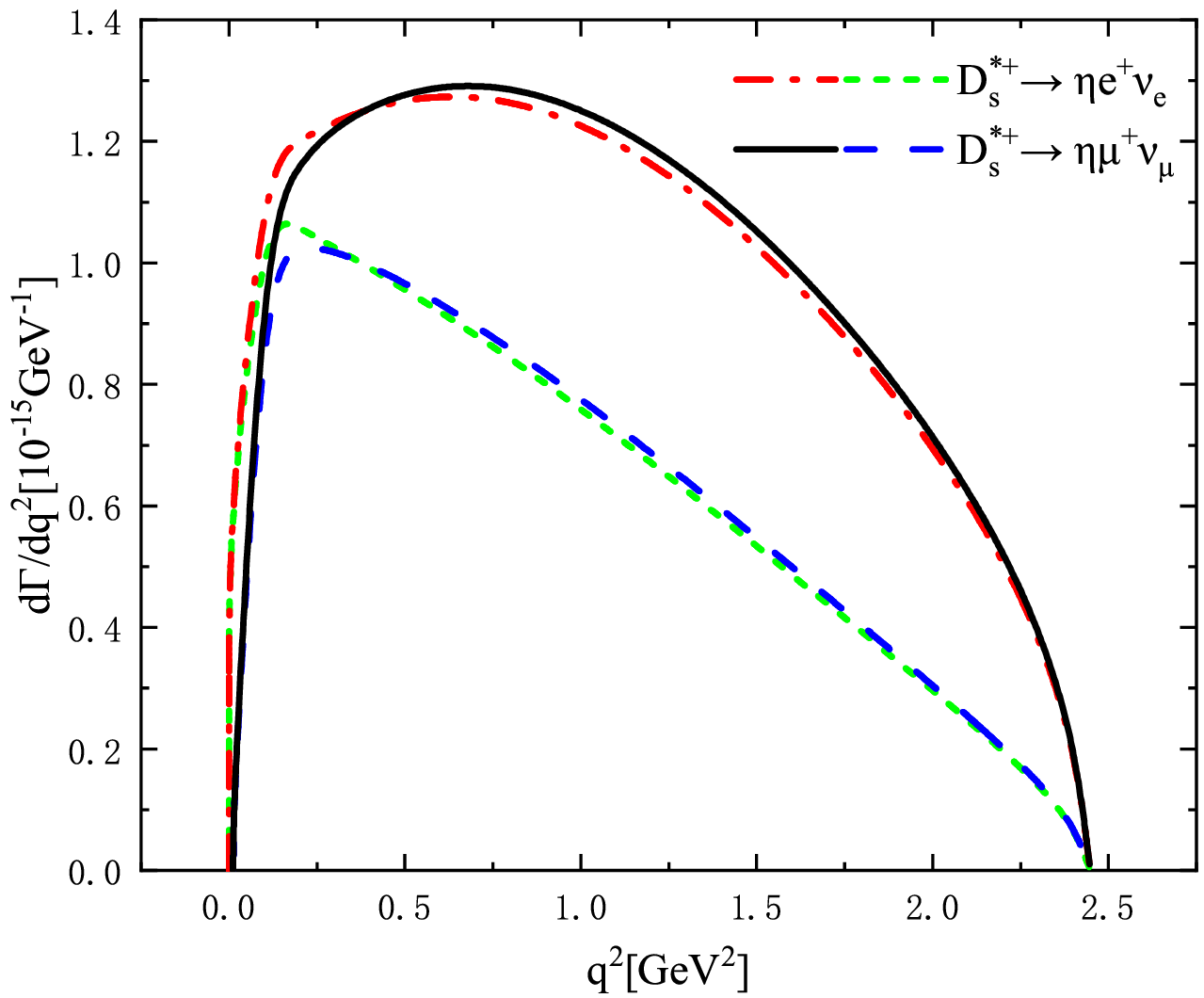}\quad}
	\subfigure[]{\includegraphics[width=0.22\textwidth]{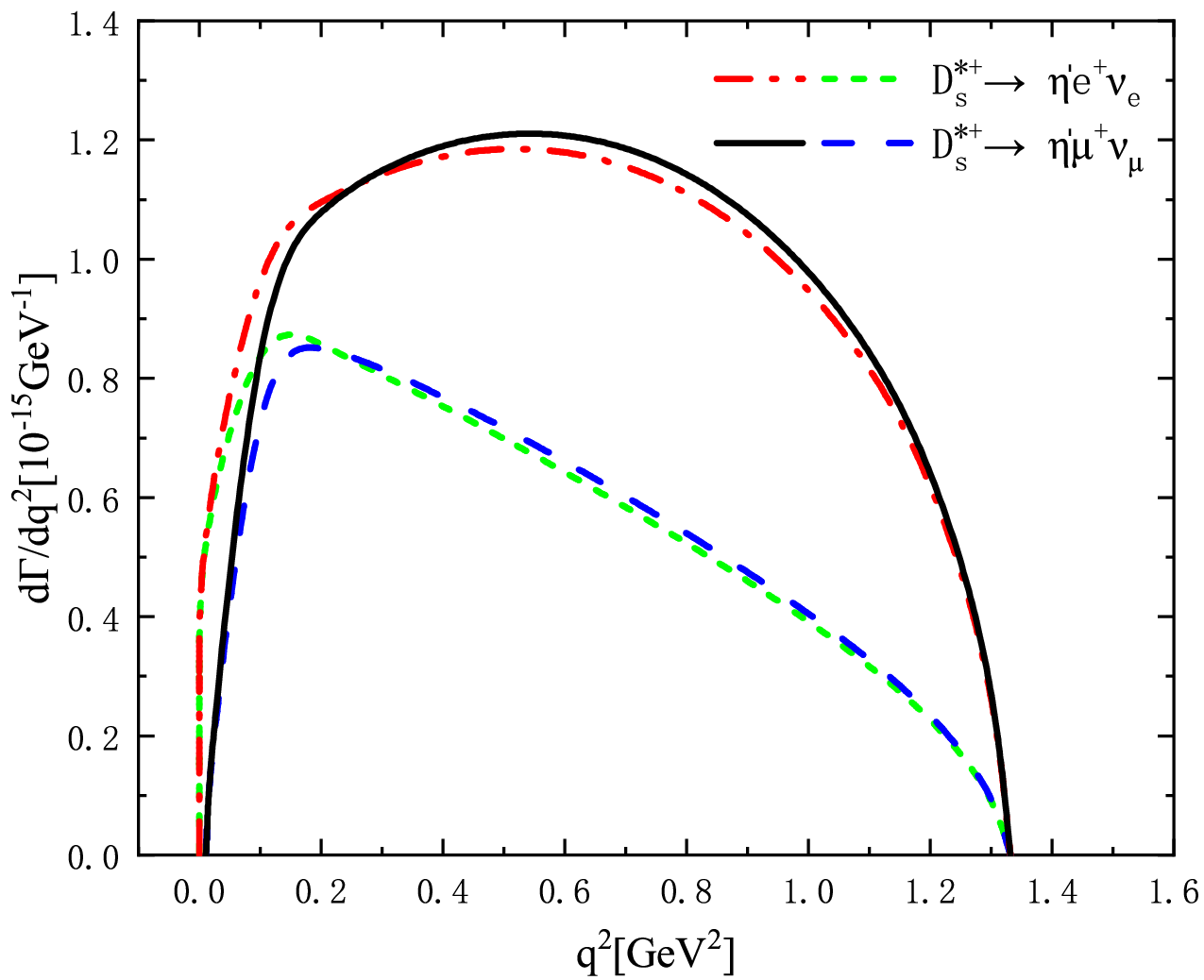}\quad}
	\subfigure[]{\includegraphics[width=0.22\textwidth]{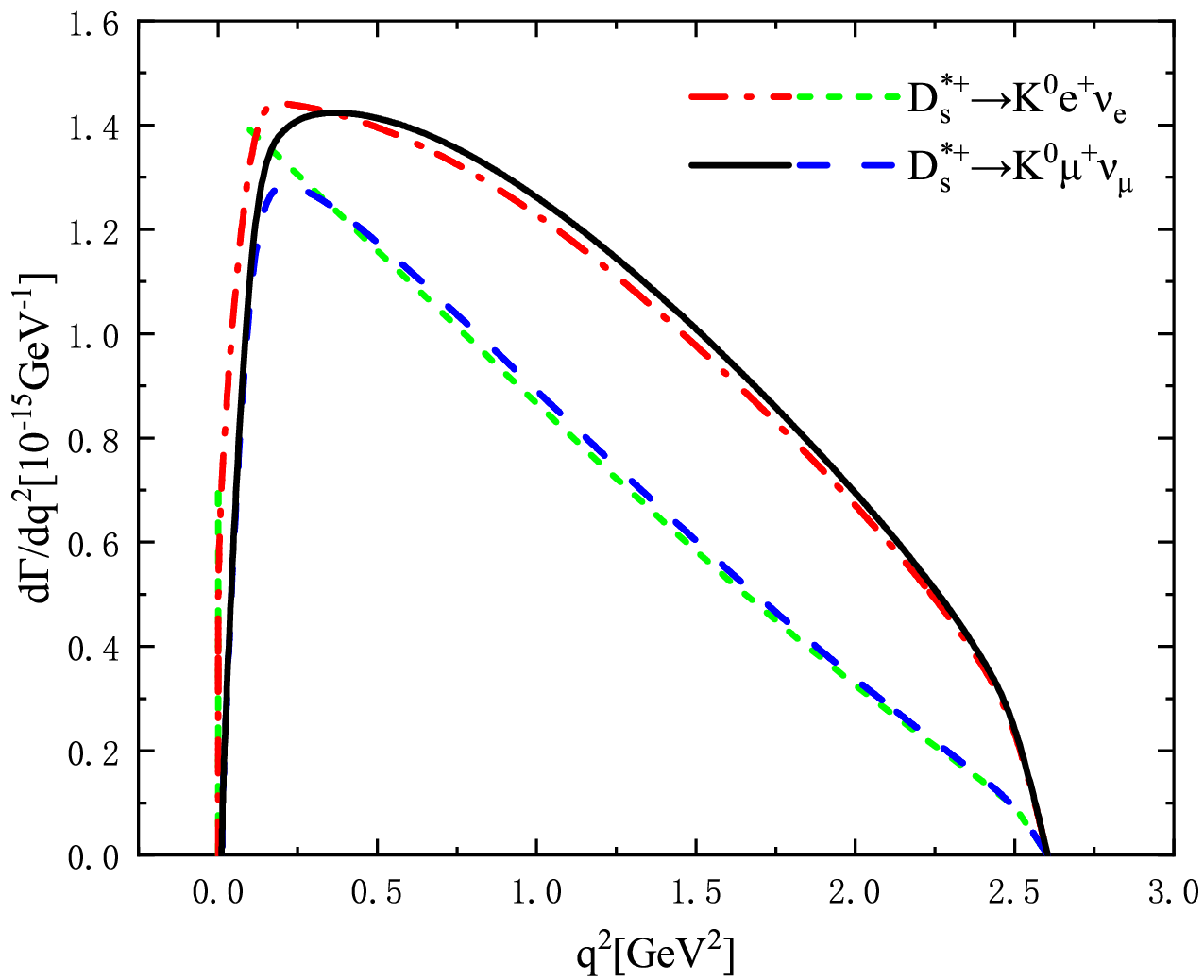}\quad}
	\caption{(color online) $q^2$ dependences of differential decay rates $d\Gamma/dq^2$ and $d\Gamma^{L}/dq^2$ for the decays $D^{*0(+)}_{(s)}\to P^{-(0)}\ell^+\nu_{\ell}$. The red and black curves represent the total polarization distributions of a given decay channel, whereas the green and blue curves correspond to the longitudinal polarization distributions of the decay channels indicated by the red and black curves, respectively.}\label{FL.}
\end{figure}

For our considered semi-leptonic decays $D^{*0(+)}_{(s)}\to P^{-(0)}\ell^+\nu_{\ell}$, we have defined the forward-backward asymmetry $A_{FB}$ in Eq. (\ref{eq:AFB}) and the longitudinal polarization fraction $f_{L}$ in Eq. (\ref{eq:fl}), to account for the impact of lepton mass and provide a more detailed physical picture. The results of these two physical observables are listed in Tables \ref{AFB} and \ref{FL}, respectively. Furthermore, in Figs. \ref{AFB.} and \ref{FL.}, we also display the $q^{2}$-dependences of the forward-backward asymmetries $A_{FB}$ and the differential decay rates $d\Gamma_{(L)}/dq^{2}$, respectively.

\begin{table}[H]
	\caption{Forward-backward asymmetries $A_{FB}$ for the decays $D^*_{(s)}\to P\ell\nu_{\ell}$, where the first error arises from the decay widths of $D_{(s)}^{*}$ and the second
		uncertainty originates from the shape parameters of the initial and final mesons.}
	\begin{center}
		\scalebox{0.85}{
			\begin{tabular}{|c|c|c|c|c|}
				\hline\hline
				Channel&$D^{*0}\rightarrow \pi ^{-}e^{+}\nu_{e}$&$D^{*0}\rightarrow \pi^{-} \mu^{+}\nu_{\mu}$&$D^{*0}\rightarrow K^{-}e^{+}\nu_{e}$&$D^{*0}\rightarrow K^{-}\mu^{+}\nu_{\mu}$\\
				\hline
				$A_{FB}$&$-0.033^{+0.003+0.001}_{-0.004-0.001}$&$-0.033^{+0.003+0.001}_{-0.004-0.001}$&$-0.148^{+0.014+0.007}_{-0.016-0.003}$&$-0.149^{+0.014+0.007}_{-0.016-0.003}$\\
				\hline
				Channel&$D^{*+}\rightarrow \pi ^{0}e^{+}\nu_{e}$&$D^{*+}\rightarrow \pi^{0} \mu^{+}\nu_{\mu}$&$D^{*+}\rightarrow \pi^{0}\tau^{+}\nu_{\tau}$&$D^{*0}\rightarrow \pi^{-}\tau^{+}\nu_{\tau}$\\
				\hline
				$A_{FB}$&$-0.031^{+0.001+0.001}_{-0.001-0.001}$&$-0.031^{+0.001+0.001}_{-0.001-0.001}$&$0.092^{+0.001+0.001}_{-0.016-0.001}$&$0.093^{+0.009+0.001}_{-0.009-0.003}$\\
				\hline
				Channel&$D^{*+}\rightarrow \bar{K}^0e^{+}\nu_{e}$&$D^{*+}\rightarrow \bar{K}^0\mu^{+}\nu_{\mu}$&$D_{s}^{*+}\rightarrow K^{0}e^{+}\nu_{e}$&$D_{s}^{*+}\rightarrow K^{0}\mu^{+}\nu_{\mu}$\\
				\hline
				$ A_{FB}$&$-0.113^{+0.002+0.005}_{-0.002-0.003}$&$-0.114^{+0.002+0.005}_{-0.003-0.003}$&$-0.161^{+0.013+0.009}_{-0.018-0.007}$&$-0.162^{+0.013+0.009}_{-0.018-0.007}$\\
				\hline
				Channel&$D^{*+}\rightarrow \eta e^{+}\nu_{e}$&$D^{*+}\rightarrow \eta \mu^{+}\nu_{\mu}$&$D^{*+}\rightarrow \eta ^{\prime}e^{+}\nu_{e}$&$D^{*+}\rightarrow \eta^{\prime}\mu^{+}\nu_{\mu}$\\
				\hline
				$A_{FB}$&$-0.184^{+0.004+0.009}_{-0.004-0.007}$&$-0.184^{+0.004+0.009}_{-0.004-0.007}$&$-0.166^{+0.004+0.008}_{-0.004-0.006}$&$-0.164^{+0.003+0.008}_{-0.004-0.006}$\\
				\hline
				Channel&$D_{s}^{*+}\rightarrow \eta e^{+}\nu_{e}$&$D_{s}^{*+}\rightarrow \eta \mu^{+}\nu_{\mu}$&$D_{s}^{*+}\rightarrow \eta ^{\prime}e^{+}\nu_{e}$&$D_{s}^{*+}\rightarrow \eta^{\prime}\mu^{+}\nu_{\mu}$\\
				\hline
				$A_{FB}$&$-0.213^{+0.018+0.010}_{-0.023-0.003}$&$-0.213^{+0.018+0.010}_{-0.023-0.003}$&$-0.200^{+0.017+0.009}_{-0.022-0.003}$&$-0.198^{+0.017+0.009}_{-0.022-0.003}$\\
				\hline\hline
			\end{tabular}\label{AFB}}
	\end{center}
\end{table}
For these decays, we observe that the values of the forward-backward asymmetries $A^{\mu}_{FB}$ and $A^e_{FB}$ are almost equal to each other. Note that the dominant contributions to the $A_{FB}$ for these decays arise from the terms proportional to $(H^2_{V,+}-H^2_{V,-})$ in Eq. (\ref{eq:btheta2}). The $A_{FB}$ of the decay $D^{*}\rightarrow \pi\tau\nu_{\tau}$ is approximately three times the size of those of the decays $D^{*}\rightarrow \pi\ell\nu_{\ell}$ with a reversed sign.
In Fig. \ref{AFB.}, the channel $D^{*0}\rightarrow \pi^{-}\tau^{+}\nu_{\tau}$ has an opposite sign for the $q^2$-dependence of the forward-backward asymmetry compared with those of the semileptonic decays $D^{*0}\rightarrow \pi^{-}\ell^{+}\nu_{\ell}$. A similar situation is observed for  the decays $D^{*+}\rightarrow \pi^{0}\tau^{+}\nu_{\tau}$ and $D^{*+}\rightarrow \pi^{0}\ell^{+}\nu_{\ell}$ not  shown in Fig. \ref{AFB.}.
The $q^2$-dependences of the
forward-backward asymmetries for the decays $D^{*+}\to P^{0}\ell^+\nu_{\ell}$ are similar to those for the decays $D^{*+}_{s}\to P^0\ell^+\nu_{\ell}$, which are not shown for simplicity.

To investigate the dependences of the polarizations on the different $q^{2}$, we divided the full energy region into two segments for each decay and calculated the longitudinal polarization fractions accordingly. Region 1 is defined as $m^{2}_{\ell} < q^{2} < \frac{(m_{D^{*}_{(s)}}-m_P)^2+m^{2}_{\ell}}{2}$, and Region 2 is $\frac{(m_{D^{*}_{(s)}}-m_P)^2+m^{2}_{\ell}}{2} < q^{2} < (m_{D^{*}_{(s)}}-m_P)^2$. The longitudinal polarization fraction in Region 1 is larger than that in Region 2 for each decay, which can be found in Table \ref{FL}.

\begin{table}[H]
	\caption{Longitudinal polarization fraction $f_{L}$ for the decays $D^{*0(+)}_{(s)}\to P^{-(0)}\ell^+\nu_{\ell}$ in Region 1 and Region 2.}
	\begin{center}
		\scalebox{0.8}{
			\begin{tabular}{cccc|cccc}
				\hline\hline
				Observables&\quad Region 1&\quad Region 2&\quad Total \quad&Observables&\quad Region 1&\quad Region 2&\quad Total\quad \\
				\hline\hline
				$f_{L}(D^{*0}\rightarrow \pi ^{-}e^{+}\nu_{e})$&$0.96$&$0.82$&$0.93$&$f_{L}(D^{*0}\rightarrow \pi^{-} \mu^{+}\nu_{\mu})$&$0.96$&$0.82$&$0.93$\\
				\hline
				$f_{L}(D^{*0}\rightarrow K^{-}e^{+}\nu_{e})$&$0.82$&$0.52$&$0.71$&$f_{L}(D^{*0}\rightarrow K^{-}\mu^{+}\nu_{\mu})$&$0.81$&$0.52$&$0.71$\\
				\hline\hline
				$f_{L}(D^{*+}\rightarrow \pi^0 e^{+}\nu_{e})$&$0.68$&$0.60$&$0.67$&$f_{L}(D^{*+}\rightarrow \pi^0 \mu^{+}\nu_{\mu})$&$0.68$&$0.60$&$0.67$\\
				\hline
				$f_{L}(D^{*+}\rightarrow \eta e^{+}\nu_{e})$&$0.77$&$0.49$&$0.67$&$f_{L}(D^{*+}\rightarrow \eta \mu^{+}\nu_{\mu})$&$0.76$&$0.49$&$0.66$\\
				\hline
				$f_{L}(D^{*+}\rightarrow \eta^{\prime}e^{+}\nu_{e})$&$0.72$&$0.44$&$0.60$&$f_{L}(D^{*+}\rightarrow \eta^{\prime}\mu^{+}\nu_{\mu})$&$0.71$&$0.44$&$0.59$\\
				\hline
				$f_{L}(D^{*+}\rightarrow \bar K^{0}e^{+}\nu_{e})$&$0.66$&$0.36$&$0.54$&$f_{L}(D^{*+}\rightarrow \bar K^{0}\mu^{+}\nu_{\mu})$&$0.65$&$0.36$&$0.53$\\
				\hline\hline
				$f_{L}(D_{s}^{*+}\rightarrow K^{0}e^{+}\nu_{e})$&$0.81$&$0.53$&$0.71$&$f_{L}(D_{s}^{*+}\rightarrow K^{0}\mu^{+}\nu_{\mu})$&$0.80$&$0.53$&$0.70$\\
				\hline
				$f_{L}(D_{s}^{*+}\rightarrow \eta e^{+}\nu_{e})$&$0.74$&$0.48$&$0.64$&$f_{L}(D_{s}^{*+}\rightarrow \eta \mu^{+}\nu_{\mu})$&$0.73$&$0.48$&$0.63$\\
				\hline
				$f_{L}(D_{s}^{*+}\rightarrow \eta ^{\prime}e^{+}\nu_{e})$&$0.70$&$0.43$&$0.58$&$f_{L}(D_{s}^{*+}\rightarrow \eta^{\prime}\mu^{+}\nu_{\mu})$&$0.69$&$0.43$&$0.57$\\
				\hline\hline
			\end{tabular}\label{FL}}
	\end{center}
\end{table}

In Table \ref{FL}, we can observe that the longitudinal polarization fractions $f_L$ of the decays $D^{*0(+)}_{(s)} \rightarrow P^{-(0)} e^+ \nu_{e}$ and $D^{*0(+)}_{(s)} \rightarrow P^{-(0)}\mu^+ \nu_{\mu}$ are close to each other, that is
\be
f_L(D^{*0(+)}_{(s)} \rightarrow P^{-(0)} e^+ \nu_{e}) \sim f_L(D^{*0(+)}_{(s)} \rightarrow P^{-(0)}\mu^+ \nu_{\mu}),
\en
which reflects the lepton flavor universality (LFU).
In each group of decays with the same intial meson, the longitudinal polarization fractions $f_L$ for the decays induced by the $c\to d \ell\nu_\ell$ transition are always larger than those of the decays induced by the $c\to s \ell\nu_\ell$.
For the decay $D^{*0(+)}\rightarrow \pi^{-(0)}\tau^{+}\nu_{\tau}$, our prediction yields the longitudinal polarization fraction $f_L$ of only 0.43, which is much smaller than those of the decays $D^{*0(+)}\to \pi^{-(0)}\ell^{+}\nu_{\ell}$ because of $m_\tau \gg m_{e, \mu}$.
A special emphasis is placed on the longitudinal polarizations for the decays $D^{*0}\rightarrow \pi ^{-}\ell^{+}\nu_{\ell}$ shown in Fig. 4(e), which are dominant and different from those for other channels. These results can be validated by the future high-luminosity experiments.
\subsection{Non-leptonic decays}
The decays rates of the nonleptonic weak decays $D^{*}_{(s)} \rightarrow PM$ with M representing a pseudoscalar meson ($\pi, K$) or a vector meson ($\rho, K^*, \phi$) can be written as \cite{ChangYang}
\begin{footnotesize}
\begin{eqnarray*}
\mathcal{B} r(D^*_{(s)} \to PM)=\frac{p_{cm}}{24\pi m_{D^{*}_{(s)}}^2\Gamma_{D^{*}_{(s)}}}|\mathcal{A}(D^*_{(s)} \to PM)|^{2},
\end{eqnarray*}
\end{footnotesize}
where $p_{cm}$ represents the three-momentum of the final meson P in the rest frame of $D^{*}_{(s)}$, and $m_{D^{*}_{(s)}}$ and $\Gamma_{D^{*}_{(s)}}$ are the mass  and the decay width of the $D^{*}_{(s)}$ meson, respectively.

\begin{table}[H]
	\caption{Branching ratios of the non-leptonic weak decays $D^{*}\to PM$. The results obtained from the naive factorization (NF) approach \cite{ChangYang} are also listed for comparison. }
	\begin{center}
		\scalebox{0.7}{
			\begin{tabular}{|c|c|c|c|c|}
				\hline\hline
				$$&$10^{-9}\times \mathcal{B}r(D^{*0}\rightarrow K^{-} \rho^{+})$&$10^{-10}\times \mathcal{B}r(D^{*0}\rightarrow K^{-} K^{*+})$&$10^{-10}\times \mathcal{B}r(D^{*0}\rightarrow K^{-} \pi^{+})$&$10^{-11}\times \mathcal{B}r(D^{*0}\rightarrow K^{-} K^{+})$\\
				\hline
				This work&$5.13^{+0.55+0.05}_{-0.49-0.16}$&$2.41^{+0.26+0.03}_{-0.23-0.09}$&$4.67^{+0.49+0.17}_{-0.45-0.29}$&$2.95^{+0.32+0.11}_{-0.28-0.19}$\\
				\hline
				\cite{ChangYang}&$2.9$&--&$7.3$&--\\
				\hline
				$$&$10^{-12}\times \mathcal{B}r(D^{*0}\rightarrow \eta \phi)$&$10^{-12}\times \mathcal{B}r(D^{*0}\rightarrow \eta^{\prime} \phi)$&$10^{-13}\times \mathcal{B}r(D^{*0}\rightarrow \eta K^{0})$&$10^{-13}\times \mathcal{B}r(D^{*0}\rightarrow \eta^{\prime} K^{0})$\\
				\hline
				This work&$8.33^{+0.95+1.4}_{-0.85-1.1}$&$1.65^{+0.17+0.20}_{-0.16-0.27}$&$3.32^{+0.35+0.11}_{-0.31-0.19}$&$1.08^{+0.12+0.02}_{-0.10-0.04}$\\
				\hline
				$$&$10^{-11}\times \mathcal{B}r(D^{*0}\rightarrow  \pi^{-} \pi^{+})$&$10^{-12}\times \mathcal{B}r(D^{*0}\rightarrow  \pi^{-} K^{+})$&$10^{-11}\times \mathcal{B}r(D^{*0}\rightarrow \pi^{0} \phi)$&$10^{-13}\times \mathcal{B}r(D^{*0}\rightarrow \pi^{0} K^{0})$\\
				\hline
				This work&$2.42^{+0.26+0.04}_{-0.23-0.10}$&$1.59^{+0.17+0.03}_{-0.15-0.07}$&$2.59^{+0.28+0.39}_{-0.25-0.35}$&$8.04^{+0.86+0.15}_{-0.77-0.34}$\\
				\hline
				$$&$10^{-9}\times \mathcal{B}r(D^{*+}\rightarrow \bar K^{0}\rho^{+})$&$10^{-10}\times \mathcal{B}r(D^{*+}\rightarrow \bar K^{0} K^{*+})$&$10^{-11}\times \mathcal{B}r(D^{*+}\rightarrow \bar K^{0}\pi^{+})$&$10^{-11}\times \mathcal{B}r(D^{*+}\rightarrow \bar K^{0} K^{+})$\\
				\hline
				This work&$3.44^{+0.08+0.03}_{-0.07-0.11}$&$1.62^{+0.04+0.02}_{-0.03-0.06}$&$9.79^{+0.22+0.52}_{-0.21-0.80}$&$2.07^{+0.04+0.08}_{-0.04-0.13}$\\
				\hline
				\cite{ChangYang}&$0.83$&--&$16$&--\\
				\hline
				$$&$10^{-11}\times \mathcal{B}r(D^{*+}\rightarrow \eta \rho^{+})$&$10^{-12}\times \mathcal{B}r(D^{*+}\rightarrow \eta K^{*+})$&$10^{-12}\times \mathcal{B}r(D^{*+}\rightarrow  \eta \pi^{+})$&$10^{-13}\times \mathcal{B}r(D^{*+}\rightarrow \eta K^{+})$\\
				\hline
				This work&$4.17^{+0.09+0.10}_{-0.09-0.09}$&$1.32^{+0.03+0.01}_{-0.03-0.03}$&$3.74^{+0.08+0.13}_{-0.08-0.22}$&$2.33^{+0.05+0.08}_{-0.05-0.14}$\\
				\hline
				$$&$10^{-11}\times \mathcal{B}r(D^{*+}\rightarrow \eta^{\prime} \rho^{+})$&$10^{-13}\times \mathcal{B}r(D^{*+}\rightarrow \eta^{\prime} K^{*+})$&$10^{-12}\times \mathcal{B}r(D^{*+}\rightarrow  \eta^{\prime} \pi^{+})$&$10^{-14}\times \mathcal{B}r(D^{*+}\rightarrow \eta^{\prime} K^{+})$\\
				\hline
				This work&$1.41^{+0.03+0.01}_{-0.03-0.03}$&$7.94^{+0.18+0.01}_{-0.17-0.15}$&$1.51^{+0.03+0.03}_{-0.03-0.06}$&$8.01^{+0.18+0.16}_{-0.17-0.35}$\\
				\hline
				$$&$10^{-11}\times \mathcal{B}r(D^{*+}\rightarrow \pi^0 \rho^{+})$&$10^{-12}\times \mathcal{B}r(D^{*+}\rightarrow \pi^0 K^{*+})$&$10^{-12}\times \mathcal{B}r(D^{*+}\rightarrow  \pi^0 \pi^{+})$&$10^{-13}\times \mathcal{B}r(D^{*+}\rightarrow \pi^0 K^{+})$\\
				\hline
				This work&$8.27^{+0.18+0.09}_{-0.17-0.08}$&$3.89^{+0.09+0.03}_{-0.08-0.04}$&$7.93^{+0.18+0.14}_{-0.17-0.33}$&$5.19^{+0.11+0.09}_{-0.11-0.22}$\\
				\hline
				$$&$10^{-11}\times \mathcal{B}r(D^{*+}\rightarrow \pi^{+} \phi)$&$10^{-12}\times \mathcal{B}r(D^{*+}\rightarrow \pi^{+} K^{0})$&$$&$$\\
				\hline
				This work&$1.40^{+0.30+0.11}_{-0.03-0.01}$&$1.06^{+0.02+0.02}_{-0.02-0.04}$&$$&$$\\
				\hline\hline
			\end{tabular}\label{Dstar}}
	\end{center}
\end{table}

\begin{table}[H]
\caption{Branching ratios of the non-leptonic decays $D^{*}_{s}\to PM$.}
\begin{center}
\scalebox{0.8}{
\begin{tabular}{|c|c|c|c|c|}
\hline\hline
$$&$10^{-7}\times \mathcal{B}r(D^{*+}_{s}\rightarrow  K^{0} \rho^{+})$&$10^{-9}\times \mathcal{B}r(D^{*+}_{s}\rightarrow  K^{0} K^{*+})$&$10^{-9}\times \mathcal{B}r(D^{*+}_{s}\rightarrow  K^{0}\pi^{+})$&$10^{-10}\times \mathcal{B}r(D^{*+}_{s}\rightarrow  K^{0} K^{+})$\\
\hline
This work&$1.17^{+0.13+0.02+0.00}_{-0.10-0.05-0.00}$&$5.33^{+0.58+0.14+0.00}_{-0.47-0.27-0.00}$&$9.38^{+1.02+0.64+0.04}_{-0.78-0.90-0.05}$&$6.07^{+0.66+0.41+0.03}_{-0.50-0.58-0.03}$\\
\hline
$$&$10^{-6}\times \mathcal{B}r(D^{*+}_{s}\rightarrow \eta \rho^{+})$&$10^{-8}\times \mathcal{B}r(D^{*+}_{s}\rightarrow \eta K^{*+})$&$10^{-8}\times \mathcal{B}r(D^{*+}_{s}\rightarrow \eta \pi^{+})$&$10^{-9}\times \mathcal{B}r(D^{*+}_{s}\rightarrow \eta K^{+})$\\
\hline
This work&$1.04^{+0.11+0.03+0.00}_{-0.09-0.05-0.00}$&$4.87^{+0.53+0.14+0.00}_{-0.40-0.27-0.00}$&$9.48^{+1.03+0.57+0.03}_{-0.78-0.85-0.02}$&$6.09^{+0.66+0.37+0.02}_{-0.50-0.55-0.01}$\\
\hline
&$10^{-7}\times \mathcal{B}r(D^{*+}_{s}\rightarrow \eta^{\prime} \rho^{+})$&$10^{-8}\times \mathcal{B}r(D^{*+}_{s}\rightarrow \eta^{\prime} K^{*+})$&$10^{-8}\times \mathcal{B}r(D^{*+}_{s}\rightarrow \eta^{\prime} \pi^{+})$&$10^{-9}\times \mathcal{B}r(D^{*+}_{s}\rightarrow \eta^{\prime} K^{+})$\\
\hline
This work&$5.90^{+0.64+0.13+0.00}_{-0.49-0.28-0.01}$&$4.65^{+0.51+0.10+0.00}_{-0.38-0.22-0.00}$&$9.05^{+0.99+0.42+0.02}_{-0.75-0.68-0.01}$&$5.14^{+0.56+0.24+0.01}_{-0.42-0.38-0.01}$\\
\hline
$$&$10^{-8}\times \mathcal{B}r(D^{*+}_{s}\rightarrow  K^{+} \phi)$&$$&$$&$$\\
\hline
This work&$2.41^{+0.26+0.06+0.01}_{-0.21-0.12-0.01}$&$$&$$&$$\\
\hline\hline
\end{tabular}\label{Dss+}}
\end{center}
\end{table}

In Tables \ref{Dstar} and \ref{Dss+}, we list the branching ratios of the non-leptonic decays $D^{*} \rightarrow PM$ and $D^{*}_s \rightarrow PM$, respectively, where the uncertainties arise from the full widths of the charmed mesons $D^{*}_{(s)}$ and the shape paramerters of the initial and final state mesons.  Numerically, we adopt the Wilson coefficients $a_{1}=1.2$ and $a_{2}=-0.5$ \cite{ChangYang}.
The following are some comments:
\begin{enumerate}
\item
The branching ratio of the decay $D^{*+}_{s}\to \eta \rho^{+}$ is four orders of magnitude larger than that of the decay $D^{*+}\to \eta \rho^{+}$. This is mainly because the decay width of the $D^{*+}$ meson $\Gamma_{D^{*+}}$ is approximately $6.8\times10^{2}$ times larger than that of $\Gamma_{D^{*+}_s}$. Furthermore, the former channel has an enhancement factor $|V_{cs}/V_{cd}|^2\approx20$ compared with the latter decay. Although the CKM matrix elements of the decay $D^{*0}\rightarrow K^{-} \rho^{+}$ are much larger than those of the decay $D^{*+}_{s}\rightarrow  K^{0} K^{*+}$, the promotion to the branching ratio from the CKM factors is almost canceled out by the large decay width $\Gamma_{D^{*0}}$. Hence, the branching ratios of these two decays are close to each other.
\item
The branching ratios of the $D^{*0}$ and $D^{*+}$ decays to the same isospin final states should be almost equal. On the other hand, the $D^{*+}$ meson decays are dynamically induced by both external and internal $W$ emission interactions, which are destructive to each other. This reason induces the hierarchical relationship, $\mathcal{B}r(D^{*+}\to \bar K^0 M)<\mathcal{B}r(D^{*0}\to K^- M)$ with $M$ referring to $\pi^+, K^+, \rho^+$ and $K^{*+}$. The branching ratios of the decays $D^*_{(s)}\to PK^*(\rho)$ are always larger than those of the corresponding decays $D^*_{(s)}\to PK(\pi)$. This is because there are three partial amplitudes for the former, whereas only the p-wave amplitude contributes to the latter. Note that our predictions for the branching ratios of the decays $D^{*0}\rightarrow K^{-} \rho^{+}$, $D^{*0}\rightarrow K^{-} \pi^{+}$, $D^{*+}\rightarrow \bar K^{0}\rho^{+}$ and $D^{*+}\rightarrow \bar K^{0}\pi^{+}$ are comptable to the results obtained by the naive factorization (NF) approach \cite{ChangYang}, which are listed in Table \ref{Dstar}.
\item
Some of the non-leptonic $D^{*+}_{s}$ decay channels are most likely to be observed in future collider experiments. In particular, the decay $D^{*+}_{s}\rightarrow \eta \rho^{+}$ has the largest branching ratio with an order of $10^{-6}$, which will be within the measurement precision and capability of the BESIII experiment in the near future.
\item
To cancel out a large part of the theoretical and experimental uncertainties and SU (3) symmetry breaking effect, it is helpful to consider the ratios of the branching ratios, such as
\begin{footnotesize}
\begin{eqnarray*}
&&R^{\eta}_{D^{*+}_s}\equiv\frac{\mathcal{B}r(D_{s}^{*+}\rightarrow \eta K^+)}{\mathcal{B}r(D_{s}^{*+}\rightarrow \eta \pi^+)}=0.064\pm{0.011},\;\;\;\;  R^{K^0}_{D^{*+}_s}\equiv\frac{\mathcal{B}r(D_{s}^{*+}\rightarrow K^0 K^+)}{\mathcal{B}r(D_{s}^{*+}\rightarrow K^0 \pi^+)}=0.065\pm0.012, \\
&&R^{\eta}_{D^{*+}}\equiv\frac{\mathcal{B}r(D^{*+}\rightarrow \eta K^+)}{\mathcal{B}r(D^{*+}\rightarrow \eta \pi^+)}=0.062^{+0.004}_{-0.006},\;\;\;\;  R^{K^-}_{D^{*0}}\equiv\frac{\mathcal{B}r(D^{*0}\rightarrow K^- K^+)}{\mathcal{B}r(D^{*0}\rightarrow K^- \pi^+)}=0.063\pm0.010,
\end{eqnarray*}
\end{footnotesize}
where the uncertainties from the transition form factors are cancelled.
\end{enumerate}

The number of potential events associated with the decays $D^{*0}\to K^-\rho^+, K^-e^+\nu_e$ and $D^{*+}_s\to \eta\rho^+, \eta e^+\nu_e$, which possess the largest branching ratios in our considered different types of decays, are listed in Table \ref{Event}. These results indicate that the study of $D^*_{(s)}$ meson weak decays through experiments is feasible.
\begin{table}[H]
\caption{Potential event numbers of the decays $D^{*0}\to K^-\rho^+, K^-e^+\nu_e$ and $D^{*+}_s\to \eta\rho^+, \eta e^+\nu_e$  in the future experiments, where the available event numbers of the $D^*_{(s)}$ have been estimated in Sec. \ref{intro}.}
\begin{center}
\scalebox{1.0}{
\begin{tabular}{c|ccccc}
\hline\hline
experiments \quad&\quad SuperKEKB&\quad STCF&\quad CEPC\cite{CEPC}&\quad FCC-ee\cite{FCCee}&\quad HL-LHC \\
\hline
$N_{D^*}$&$2\times10^{10}$&$8\times10^{10}$&$10^{11}$&$10^{12}$&$2\times10^{14}$\\
\hline
$N_{D^{*0}\to K^-\rho^+}$&$102$&$410$&$513$&$5.13\times 10^3$&$1.03\times10^6$\\
$N_{D^{*0}\to K^-e^+\nu_e}$&$158$&$634$&$793$&$7.93\times 10^3$&$1.59\times10^6$\\
\hline\hline
$N_{D^{*+}_s}$&$5.5\times10^{9}$&$10^{10}$&$1.3\times10^{10}$&$6.6\times10^{10}$&$4\times10^{13}$\\
\hline
$N_{D^{*+}_s\to \eta\rho^+}$&$5.72\times10^3$&$1.04\times10^4$&$1.35\times10^4$&$6.86\times10^4$&$3.94\times10^7$\\
$N_{D^{*+}_s \to \eta e^+\nu_e}$&$8.03\times 10^3$&$1.46\times 10^4$&$1.89\times 10^4$&$9.63\times 10^4$&$5.84\times 10^7$\\
\hline\hline
\end{tabular}\label{Event}}
\end{center}
\end{table}

\section{Summary}\label{sum}
Although the $D^*$ and $D^*_{s}$ mesons were discovered more than 45 years ago,  the information about their properties is still relatively limited, in particular,  the measurements of the $D^{*}_{(s)}$ weak decays are still unavailable from experiments.  Inspired by the recent advances and future prospects for the study of $D^*_{(s)}$ mesons in the collider experiments, we explore both the semi-leptonic and non-leptonic $D^*_{(s)}$ weak decays. Combining the helicity amplitudes with the form factors of the transitions $D^*_{(s)}\to \pi, K, \eta_{q,s}$ obtained from the CLFQM, the branching ratios of the corresponding $D^*_{(s)}$ weak decays are calculated. The $D^*$ weak decays can be measured by future collider experiments with the ability to observe the branching ratio with an order of $10^{-9}$, such as the FCC-ee and HL-LHC. Compared with the $D^*$ meson, the $D^*_s$ meson weak decays are relatively easier to observe in the future experiments. For example, the branching ratios of the decays $D_{s}^{*+}\to\eta \ell^{+}\nu_{\ell}$ and $D^{*+}_{s}\to \eta\rho^{+}$ can reach up to the order of $10^{-6}$, which
correspond to tens of thousands of events in the $e^+e^-$ collider experiments and tens of millions of events at the HL-LHC. Furthermore, to provide a more detailed physical picture for our considered decays, the longitudinal polarization fraction $f_{L}$ and forward-backward asymmetry $A_{FB}$ are calculated.
\section*{Acknowledgment}
This work is partly supported by the National Natural Science
Foundation of China under Grant No. 11347030, by the Program of
Science and Technology Innovation Talents in Universities of Henan
Province 14HASTIT037, and the Natural Science Foundation of Henan
Province under grant no. 232300420116, 252300421302. We would like to thank Prof. Junfeng Sun for helpful discussions.
\appendix
\section{SOME USEFUL FORMULAS }
The $p_1$ and $p_2$ are the on-mass-shell light-front momenta,
\be
\tilde{p}=(p^+, p_\perp), \hspace{0.3cm} p_\perp =(p^1, p^2), \hspace{0.3cm} p^-=\frac{m^2+p^2_\perp}{p^+}
\en
The light-front relative momentum variables $(x, p_\perp)$ defined by
\be
p_{1}^{+}=x_{1} P^{+}, \quad p_{2}^{+}=x_{2} P^{+}, \quad x_{1}+x_{2}=1, \\
p_{1 \perp}=x_{1} P_{\perp}+p_{\perp}, \quad p_{2 \perp}=x_{2} P_{\perp}-p_{\perp},
\en

Some specific rules are provided under the $p^-$ integration. When integrating, it is important to include the zero-mode contribution for proper integration in the CLFQM. Specifically, we follow the rules outlined in Ref.\cite{ChengHwang}
\be
\hat{p}_{1 \mu}^{\prime} &\doteq &   P_{\mu}
A_{1}^{(1)}+q_{\mu} A_{2}^{(1)},\\
\hat{p}_{1 \mu}^{\prime}
\hat{p}_{1 \nu}^{\prime}  &\doteq & g_{\mu \nu} A_{1}^{(2)} +P_{\mu}
P_{\nu} A_{2}^{(2)}+\left(P_{\mu} q_{\nu}+q_{\mu} P_{\nu}\right)
A_{3}^{(2)}+q_{\mu} q_{\nu} A_{4}^{(2)},\\
Z_{2}&=&\hat{N}_{1}^{\prime}+m_{1}^{\prime 2}-m_{2}^{2}+\left(1-2
x_{1}\right) M^{\prime 2} +\left(q^{2}+q \cdot P\right)
\frac{p_{\perp}^{\prime} \cdot q_{\perp}}{q^{2}},\\
\hat{p}_{1 \mu}^{\prime} \hat{N}_{2} & \rightarrow & q_{\mu}\left[A_{2}^{(1)} Z_{2}+\frac{q \cdot P}{q^{2}} A_{1}^{(2)}\right],\\
\hat{p}_{1 \mu}^{\prime} \hat{p}_{1 \nu}^{\prime} \hat{N}_{2} & \rightarrow &g_{\mu \nu} A_{1}^{(2)} Z_{2}+q_{\mu} q_{\nu}\left[A_{4}^{(2)} Z_{2}+2 \frac{q \cdot P}{q^{2}} A_{2}^{(1)} A_{1}^{(2)}\right],\\
A_{1}^{(1)}&=&\frac{x_{1}}{2}, \quad A_{2}^{(1)}=
A_{1}^{(1)}-\frac{p_{\perp}^{\prime} \cdot q_{\perp}}{q^{2}},\quad A_{3}^{(2)}=A_{1}^{(1)} A_{2}^{(1)},\\
A_{4}^{(2)}&=&\left(A_{2}^{(1)}\right)^{2}-\frac{1}{q^{2}}A_{1}^{(2)},\quad A_{1}^{(2)}=-p_{\perp}^{\prime 2}-\frac{\left(p_{\perp}^{\prime}
\cdot q_{\perp}\right)^{2}}{q^{2}}, \quad A_{2}^{(2)}=\left(A_{1}^{(1)}\right)^{2}.
\en
\section{The amplitudes for $D^{*}_{(s)} \rightarrow PM$ decays}
\be
S_{\mu \nu}^{D^{*}_{(s)} P}&=&\left(S_{V}^{D^{*}_{(s)} P}-S_{A}^{D^{*}_{(s)} P}\right)_{\mu \nu}\non
&=&\operatorname{Tr}\left[\left(\gamma_{\nu}-\frac{1}{W_{V}^{\prime \prime}}\left(p_{1}^{\prime \prime}-p_{2}\right)_{\nu}\right)\left(p_{1}^{\prime \prime}
+m_{1}^{\prime \prime}\right)\left(\gamma_{\mu}-\gamma_{\mu} \gamma_{5}\right)\left(\not p_{1}^{\prime}+m_{1}^{\prime}\right) \gamma_{5}\left(-\not p_{2}
+m_{2}\right)\right] \non
&=&-2 i \epsilon_{\mu \nu \alpha \beta}\left\{p_{1}^{\prime \alpha} P^{\beta}\left(m_{1}^{\prime \prime}-m_{1}^{\prime}\right)
+p_{1}^{\prime \alpha} q^{\beta}\left(m_{1}^{\prime \prime}+m_{1}^{\prime}-2 m_{2}\right)+q^{\alpha} P^{\beta} m_{1}^{\prime}\right\} \non
&&+\frac{1}{W_{V}^{\prime \prime}}\left(4 p_{1 \nu}^{\prime}-3 q_{\nu}-P_{\nu}\right) i \epsilon_{\mu \alpha \beta \rho} p_{1}^{\prime \alpha} q^{\beta} P^{\rho}\non &&
+2 g_{\mu \nu}\left\{m_{2}\left(q^{2}-N_{1}^{\prime}-N_{1}^{\prime \prime}-m_{1}^{\prime 2}-m_{1}^{\prime \prime 2}\right)
-m_{1}^{\prime}\left(M^{\prime \prime 2}-N_{1}^{\prime \prime}-N_{2}-m_{1}^{\prime \prime 2}-m_{2}^{2}\right)\right.\non
&&\left.-m_{1}^{\prime \prime}\left(M^{\prime 2}-N_{1}^{\prime}-N_{2}-m_{1}^{\prime 2}-m_{2}^{2}\right)
-2 m_{1}^{\prime} m_{1}^{\prime \prime} m_{2}\right\} \non &&
+8 p_{1 \mu}^{\prime} p_{1 \nu}^{\prime}\left(m_{2}-m_{1}^{\prime}\right)-2\left(P_{\mu} q_{\nu}
+q_{\mu} P_{\nu}+2 q_{\mu} q_{\nu}\right) m_{1}^{\prime}+2 p_{1 \mu}^{\prime} P_{\nu}\left(m_{1}^{\prime}-m_{1}^{\prime \prime}\right)\non &&
+2 p_{1 \mu}^{\prime} q_{\nu}\left(3 m_{1}^{\prime}-m_{1}^{\prime \prime}-2 m_{2}\right)
+2 P_{\mu} p_{1 \nu}^{\prime}\left(m_{1}^{\prime}+m_{1}^{\prime \prime}\right)+2 q_{\mu} p_{1 \nu}^{\prime}\left(3 m_{1}^{\prime}+m_{1}^{\prime \prime}-2 m_{2}\right)\non &&
+\frac{1}{2 W_{V}^{\prime \prime}}\left(4 p_{1 \nu}^{\prime}-3 q_{\nu}-P_{\nu}\right)\left\{2 p_{1 \mu}^{\prime}\left[M^{\prime 2}
+M^{\prime \prime 2}-q^{2}-2 N_{2}+2\left(m_{1}^{\prime}-m_{2}\right)\left(m_{1}^{\prime \prime}+m_{2}\right)\right]\right.\non&&
+q_{\mu}\left[q^{2}-2 M^{\prime 2}+N_{1}^{\prime}-N_{1}^{\prime \prime}+2 N_{2}-\left(m_{1}+m_{1}^{\prime \prime}\right)^{2}+2\left(m_{1}^{\prime}-m_{2}\right)^{2}\right]\non&&
\left.+P_{\mu}\left[q^{2}-N_{1}^{\prime}-N_{1}^{\prime \prime}-\left(m_{1}^{\prime}+m_{1}^{\prime \prime}\right)^{2}\right]\right\} .
\label{sptov}
\en

The following formulas are the analytical expressions of the form factors of transitions $D^{*}_{(s)} \rightarrow P$ in the CLFQM:

\begin{footnotesize}
\begin{eqnarray}
V^{D^{*}_{(s)} P}(q^{2})&=&\frac{N_{c}(M^{'}+M^{''})}{16 \pi^{3}} \int d x_{2} d^{2} p_{\perp}^{\prime} \frac{2 h_{D^{*}_{(s)}}^{\prime}
 h_{P}^{\prime \prime}}{x_{2} \hat{N}_{1}^{\prime} \hat{N}_{1}^{\prime \prime}}\left\{x_{2} m_{1}^{\prime}
 +x_{1} m_{2}+\left(m_{1}^{\prime}-m_{1}^{\prime \prime}\right) \frac{p_{\perp}^{\prime} \cdot q_{\perp}}{q^{2}}\right.\non &&\left.
 +\frac{2}{w_{D^{*}_{(s)}}^{\prime \prime}}\left[p_{\perp}^{\prime 2}+\frac{\left(p_{\perp}^{\prime} \cdot q_{\perp}\right)^{2}}{q^{2}}\right]\right\},\\
 A_0^{D^{*}_{(s)} P}(q^{2})&=& \frac{M^{'}+M^{''}}{2M^{''}}A_1^{D^{*}_{(s)} P}(q^{2})-\frac{M^{'}-M^{''}}{2M^{''}}A_2^{D^{*}_{(s)} P}(q^{2})-\frac{q^2}{2M^{''}}\frac{N_{c}}{16 \pi^{3}} \int d x_{2} d^{2} p_{\perp}^{\prime} \frac{h_{D^{*}_{(s)}}^{\prime} h_{P}^{\prime \prime}}{x_{2} \hat{N}_{1}^{\prime}
\hat{N}_{1}^{\prime \prime}}\left\{2\left(2 x_{1}-3\right)\right.\non &&\left.\left(x_{2} m_{1}^{\prime}+x_{1} m_{2}\right)-8\left(m_{1}^{\prime}-m_{2}\right)
\times\left[\frac{p_{\perp}^{\prime 2}}{q^{2}}
+2 \frac{\left(p_{\perp}^{\prime} \cdot q_{\perp}\right)^{2}}{q^{4}}\right]-\left[\left(14-12 x_{1}\right) m_{1}^{\prime}\right.\right. \non &&\left.\left.-2 m_{1}^{\prime \prime}-\left(8-12 x_{1}\right) m_{2}\right] \frac{p_{\perp}^{\prime} \cdot q_{\perp}}{q^{2}}
+\frac{4}{w_{D^{*}_{(s)}}^{\prime \prime}}\left(\left[M^{\prime 2}+M^{\prime \prime 2}-q^{2}+2\left(m_{1}^{\prime}-m_{2}\right)\left(m_{1}^{\prime \prime}
+m_{2}\right)\right]\right.\right.\non &&\left.\left.\times\left(A_{3}^{(2)}+A_{4}^{(2)}-A_{2}^{(1)}\right)
+Z_{2}\left(3 A_{2}^{(1)}-2 A_{4}^{(2)}-1\right)+\frac{1}{2}\left[x_{1}\left(q^{2}+q \cdot P\right)
-2 M^{\prime 2}-2 p_{\perp}^{\prime} \cdot q_{\perp}\right.\right.\right.\non &&\left.\left.\left.-2 m_{1}^{\prime}\left(m_{1}^{\prime \prime}+m_{2}\right)
-2 m_{2}\left(m_{1}^{\prime}-m_{2}\right)\right]\left(A_{1}^{(1)}+A_{2}^{(1)}-1\right) q \cdot P\left[\frac{p_{\perp}^{\prime 2}}{q^{2}}
+\frac{\left(p_{\perp}^{\prime} \cdot q_{\perp}\right)^{2}}{q^{4}}\right]\right.\right.\non &&\left.\left.\times\left(4 A_{2}^{(1)}-3\right)\right)\right\},
\end{eqnarray}
\end{footnotesize}

\begin{footnotesize}
\begin{eqnarray}
A_1^{D^{*}_{(s)} P}(q^{2})&=& -\frac{1}{M^{'}+M^{''}}\frac{N_{c}}{16 \pi^{3}} \int d x_{2} d^{2} p_{\perp}^{\prime} \frac{h_{D^{*}_{(s)}}^{\prime} h_{P}^{\prime \prime}}{x_{2}
\hat{N}_{1}^{\prime}
\hat{N}_{1}^{\prime \prime}}\left\{2 x_{1}\left(m_{2}-m_{1}^{\prime}\right)\left(M_{0}^{\prime 2}+M_{0}^{\prime \prime 2}\right)
-4 x_{1} m_{1}^{\prime \prime} M_{0}^{\prime 2}\right.\non
&&\left.+2 x_{2} m_{1}^{\prime} q \cdot P+2 m_{2} q^{2}-2 x_{1} m_{2}\left(M^{\prime 2}+M^{\prime \prime 2}\right)+2\left(m_{1}^{\prime}-m_{2}\right)\left(m_{1}^{\prime}
+m_{1}^{\prime \prime}\right)^{2}+8\left(m_{1}^{\prime}-m_{2}\right) \right.\non &&
\left. \times\left[p_{\perp}^{\prime 2}+\frac{\left(p_{\perp}^{\prime}
\cdot q_{\perp}\right)^{2}}{q^{2}}\right]+2\left(m_{1}^{\prime}+m_{1}^{\prime \prime}\right)\left(q^{2}+q \cdot P\right) \frac{p_{\perp}^{\prime} \cdot q_{\perp}}{q^{2}}
-4 \frac{q^{2} p_{\perp}^{\prime 2}+\left(p_{\perp}^{\prime} \cdot q_{\perp}\right)^{2}}{q^{2} w_{D^{*}_{(s)}}^{\prime \prime}}
\right.\non && \left.\times\left[2 x_{1}\left(M^{\prime 2}+M_{0}^{\prime 2}\right)-q^{2}-q \cdot P-2\left(q^{2}+q \cdot P\right) \frac{p_{\perp}^{\prime} \cdot q_{\perp}}{q^{2}}-2\left(m_{1}^{\prime}-m_{1}^{\prime \prime}\right)\left(m_{1}^{\prime}-m_{2}\right)\right]\right\},\;\;\;\;\;\\
A_2^{D^{*}_{(s)} P}(q^{2})&=& \frac{N_{c}(M^{'}+M^{''})}{16 \pi^{3}} \int d x_{2} d^{2} p_{\perp}^{\prime} \frac{2 h_{D^{*}_{(s)}}^{\prime} h_{P}^{\prime \prime}}{x_{2} \hat{N}_{1}^{\prime}
\hat{N}_{1}^{\prime \prime}}\left\{\left(x_{1}-x_{2}\right)\left(x_{2} m_{1}^{\prime}+x_{1} m_{2}\right)-\frac{p_{\perp}^{\prime} \cdot q_{\perp}}{q^{2}}\left[2 x_{1} m_{2}
+m_{1}^{\prime \prime} \right.\right.\non &&
\left.\left.+\left(x_{2}-x_{1}\right) m_{1}^{\prime}\right]-2 \frac{x_{2} q^{2}+p_{\perp}^{\prime} \cdot q_{\perp}}{x_{2} q^{2} w_{D^{*}_{(s)}}^{\prime \prime}}\left[p_{\perp}^{\prime} \cdot p_{\perp}^{\prime \prime}
+\left(x_{1} m_{2}+x_{2} m_{1}^{\prime}\right)\left(x_{1} m_{2}-x_{2} m_{1}^{\prime \prime}\right)\right]\right\}.\\
A^{D^{*}_{(s)}P}_3(0)&=&A^{D^{*}_{(s)}P}_0(0) \hspace{0.3cm} \text{and} \hspace{0.3cm} A^{D^{*}_{(s)}P}_3(q^2)=\frac{m_{D_{(s)}^*}+m_{P}}{2m_{P}}A_{1}^{D_{(s)}^* P}(q^2)-\frac{m_{D_{(s)}^*}-m_{P}}{2m_{P}}A_{2}^{D_{(s)}^* P}(q^2).
\end{eqnarray}
\end{footnotesize}


\end{document}